\documentclass[a4paper,11pt]{article}
\pdfoutput=0 

\usepackage{jheppub} 

\usepackage[T1]{fontenc} 
\usepackage{slashed}

\newcommand{\gsim}{\;\rlap{\lower 3.5 pt \hbox{$\mathchar \sim$}} \raise 1pt
\hbox {$>$}\;}
\newcommand{\lsim}{\;\rlap{\lower 3.5 pt \hbox{$\mathchar \sim$}} \raise 1pt
\hbox {$<$}\;}

\title{\boldmath
Higgs Boson Production and Quark Scattering Amplitudes at High Energy through the
Next-to-Next-to-Leading Power in Quark Mass}

\preprint{ALBERTA-THY-32-21}

\author[a,b]{Tao Liu,}
\author[c]{Sneh Modi,}
\author[c]{Alexander A. Penin}

\affiliation[a]{Institute of High Energy Physics,  Chinese Academy of Sciences, Beijing 100049, China}
\affiliation[b]{University of Chinese Academy of Sciences, Beijing 100049, China}
\affiliation[c]{Department of Physics, University of Alberta,
Edmonton AB T6G 2J1, Canada
}

\emailAdd{liutao86@ihep.ac.cn}
\emailAdd{smodi@ualberta.ca}
\emailAdd{penin@ualberta.ca}

\abstract{We study  the amplitudes  of the quark scattering by
an external electromagnetic field  and of the light quark
mediated  Higgs boson production via gluon fusion in the
high-energy limit. The asymptotic behavior of the quark form
factors is obtained in the double-logarithmic approximation to
all orders in strong coupling constant through ${\cal
O}(m_q^3)$ in the small quark mass expansion and the asymptotic
formula is given in a closed analytic form. In the case of
the  two-gluon Higgs boson form factor we obtain  a complete
analytic result for the  three-loop ${\cal O}(m_q^3)$
double-logarithmic term while the all-order analysis  is
performed  in the large-$N_c$ limit of QCD and for the abelian
gauge group. An estimate of the high-order high-power light
quark mass effect in the Higgs boson production and decay is
given.}

\begin{document}
\maketitle
\flushbottom

\section{Introduction}

Quantum corrections are known to significantly alter the
high-energy properties of the gauge theory scattering
amplitudes. The asymptotic behavior of the amplitudes which are
not suppressed by the ratio of a characteristic infrared scale
to the process energy is governed by the ``Sudakov'' radiative
corrections enhanced by the second power of the large logarithm
of the scale ratio per each power of the coupling constant.
Sudakov  logarithms   exponentiate and result in a strong
universal suppression of the scattering amplitudes in the limit
when all the kinematic invariants of the process are large
\cite{Sudakov:1954sw,Frenkel:1976bj,Smilga:1979uj,Mueller:1979ih,
Collins:1980ih,Sen:1981sd,Sterman:1986aj,Korchemsky:1988hd,
Korchemsky:1988pn}. The structure of the power suppressed
logarithmically enhanced contributions is by far  more complex
and the corresponding renormalization group analysis poses a
serious  challenge to the modern effective field theory. One of
the important problems in  this category is the analysis of the
scattering amplitudes involving massive particle in the limit
of small mass or high energy. The mass effects on  the
leading-power contributions have been extensively studied in
the context of the high-order electroweak and QED radiative
corrections
\cite{Kuhn:1999nn,Kuhn:2001hz,Feucht:2004rp,Jantzen:2005az,
Penin:2005eh,Penin:2005kf,Bonciani:2007eh,Bonciani:2008ep,
Kuhn:2007ca,Kuhn:2011mh,Penin:2011aa}. The next-to-leading
power contributions for a number of  key processes in QED and
QCD have been analysed in the leading (double)
\cite{Gorshkov:1966ht,Kotsky:1997rq,Penin:2014msa,
Melnikov:2016emg,Penin:2016wiw,Liu:2017axv,Liu:2017vkm,
Liu:2018czl,Liu:2019oav,Wang:2019mym} and the next-to-leading
logarithmic approximation
\cite{Anastasiou:2020vkr,Liu:2020tzd,Liu:2020wbn}.\footnote{The
next-to-leading power logarithmic contributions corrections
have also been recently discussed  in many different
incarnations
\cite{Ferroglia:2009ep,Laenen:2010uz,Becher:2013iya,
deFlorian:2014vta,Anastasiou:2014lda,
Boughezal:2018mvf,Bruser:2018jnc,Moult:2018jjd,Ebert:2018lzn,
Beneke:2018gvs,Engel:2018fsb,Ebert:2018gsn,
Penin:2019xql,Beneke:2019mua}.}

In the processes with massive fermions already at the
next-to-leading power  the origin of the logarithmic
corrections and the asymptotic behavior of the amplitudes
drastically differ from the leading-power Sudakov case. The
double-logarithmic terms in this case are related to the effect
of the eikonal (color) charge nonconservation in the process
with soft fermion exchange and result in asymptotic exponential
enhancement for a wide class of amplitudes and in a breakdown
of a formal power counting
\cite{Penin:2014msa,Liu:2017vkm,Liu:2018czl}. Thus, it is of a
primary theoretical interest to get insight into the asymptotic
behavior of the  next-to-next-to-leading power contributions
and determine whether any qualitatively new phenomenon appears
in this order. The renormalization group analysis has not yet
been extended   beyond the next-to-leading power for any kind
of power  corrections to the high-energy processes.  In this
paper  we present for the first time such an analysis of the
simplest but fundamental and phenomenologically important
amplitudes of the quark scattering in an external
electromagnetic field and of the light quark mediated Higgs boson
production in gluon fusion. The results of the  analysis are
used to get  a quantitative estimate of the accuracy of the
fixed-order calculations
\cite{Czakon:2020vql,Niggetiedt:2020sbf} and the calculations
based on the small-mass expansion
\cite{Melnikov:2016qoc,Lindert:2017pky} of the light quark
contribution to the Higgs boson production and decays.

The paper is organized as follows. In the next section we
discuss the scattering of a massive quark by an external
electromagnetic field in the limit of large momentum transfer,
recall the main features of the double-logarithmic result for
the next-to-leading power contribution and extend the analysis
to the ${\cal O}(m_q^3)$ amplitude. In Sect.~\ref{sec::3} we
discuss the amplitude of the Higgs boson production at ${\cal
O}(m_q^3)$, derive  the analytic result for the three-loop
double-logarithmic term and extend it to  all orders in the
large-$N_c$ limit and in the case of  the abelian gauge group.
Sect.~\ref{sec::4} is our summary.

\section{Quark scattering by electromagnetic field}
\label{sec::2}

The  amplitude  ${\cal F}$  of a quark scattering in an
external field can be parameterized in the standard way  by the
Dirac and Pauli form factors
\begin{equation}
{\cal F}=e_q\bar{\psi}(p_1)\left(\gamma_\mu F_1
+{i\sigma_{\mu\nu}q^\nu \over 2m_q}
F_2\right)\psi(p_2)\,,
\label{eq::Dirac}
\end{equation}
where $e_q$ is the quark charge. For  on-shell quark
$p_1^2=p_2^2=m_q^2$ and the large Euclidean momentum transfer
$Q^2=-(p_2-p_1)^2$ when the ratio $\rho\equiv {m_q^2/Q^2}$ is
positive and small the form factors can be expanded in an
asymptotic series
\begin{equation}
F_i=Z_{q}^2\sum_{n=0}^\infty \rho^n F^{(n)}_i\,,
\label{eq::Fiseries}
\end{equation}
where the universal Sudakov factor for the external on-shell
quark lines which incorporates all the infrared divergencies of
the amplitude. In dimensional regularization with
$d=4-2\varepsilon$ in the double-logarithmic approximation it
reads
\begin{equation}
 Z_{q}^2=\exp\left[-C_F\left({\alpha_s\over 2\pi}
 {\ln\rho\over \varepsilon}+x\right)\right],
\label{eq::Zq}
\end{equation}
where $x={\alpha_s\over 4\pi}\ln^2\!\rho$ is the
double-logarithmic variable and  $C_F=(N_c^2-1)/(2N_c)$ for the
$SU(N_c)$ color group. The infrared finite coefficients
$F_i^{(n)}$ in a given order of perturbation theory depends on
$\rho$ only logarithmically, and in the double-logarithmic
approximation are functions of $x$. Due to factorization of
Sudakov logarithms into $Z_{q}^2$ these coefficients include
only non-Sudakov double logarithms and the leading-power Dirac
form factor  with the logarithmic accuracy is just
$F_1^{(0)}=1$. At the same time the Pauli form factor describe
the scattering with a flip of the quark chirality  and
therefore has to vanish in the high-energy or small-mass limit
{\it i.e.} $F_2^{(0)}=0$.

The next-to-leading power double-logarithmic contribution to
the Dirac form factor is generated  by the soft quark pair
exchange  and starts with two loops. The corresponding
coefficient reads
\cite{Liu:2017vkm,Liu:2018czl}
\begin{equation}
F_1^{(1)}={C_F(C_A-2C_F)\over 6}x^2f\left(-z\right)\,,
\label{eq::F1result}
\end{equation}
where $z =(C_A-C_F)x$, $C_A=N_c$ and the function $f$  has
the following  integral representation
\begin{equation}
f(z)=12\int_0^1{\rm d}\eta_1\int_{\eta_1}^{1}{\rm d}\eta_2
\int_0^{1-\eta_2}{\rm d}\xi_2\int_{\xi_2}^{1-\eta_1}
 {\rm d}\xi_1\,e^{2z\eta_1(\xi_1-\xi_2)} e^{2z\xi_2(\eta_2-\eta_1)}\,.
\label{eq::f}
\end{equation}
Due to the $1/m_q$ factor in the definition of $F_2$ the
coefficient $F_2^{(1)}$ corresponds to the  ${\cal O}(m_q)$
scattering amplitude. However, as we will see in the
double-logarithmic approximation $F_2^{(1)}=0$ and the Pauli
form factor starts to contribute at ${\cal O}(m_q^3)$,  {\it
i.e.} at the next-to-next-to-leading power in small-mass
expansion.

\begin{figure}[t]
\begin{center}
\begin{tabular}{cccc}
\includegraphics[width=2.4cm]{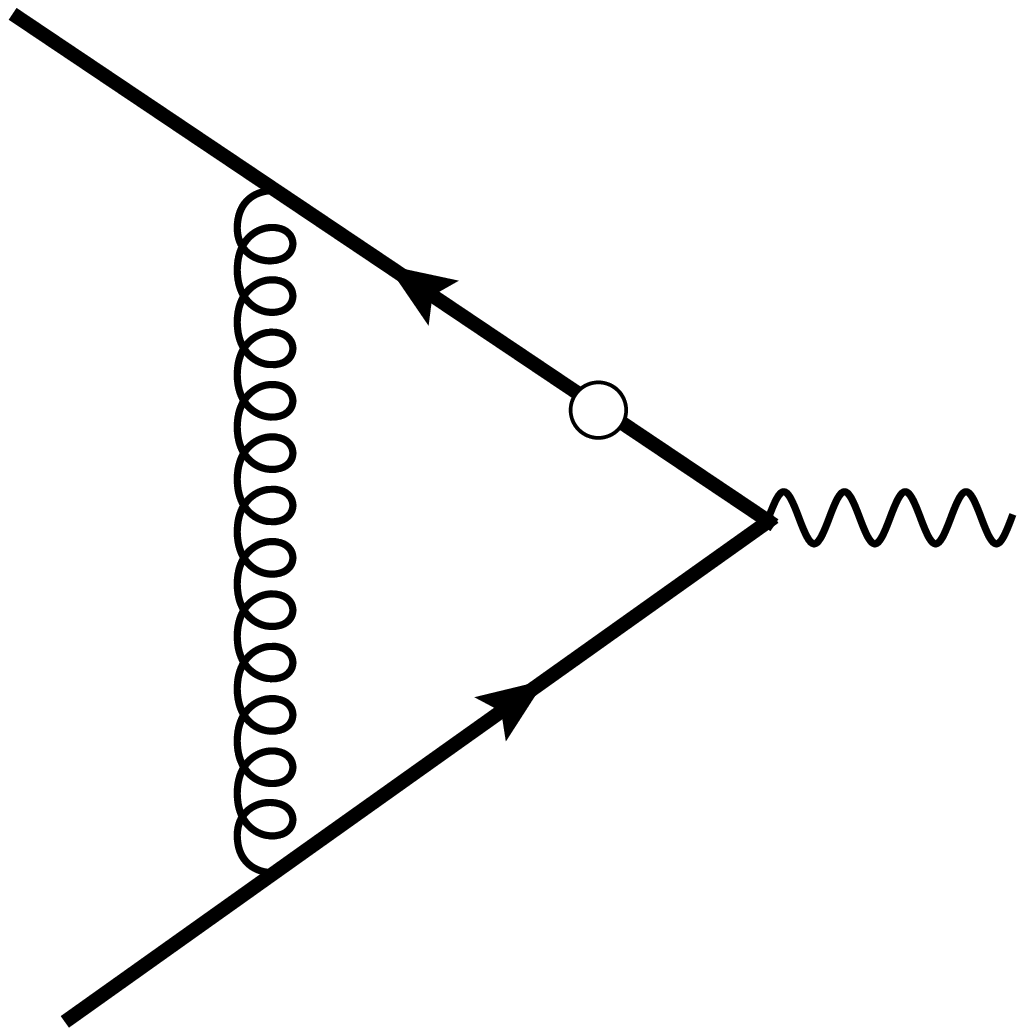}&
\hspace*{03mm}\includegraphics[width=2.4cm]{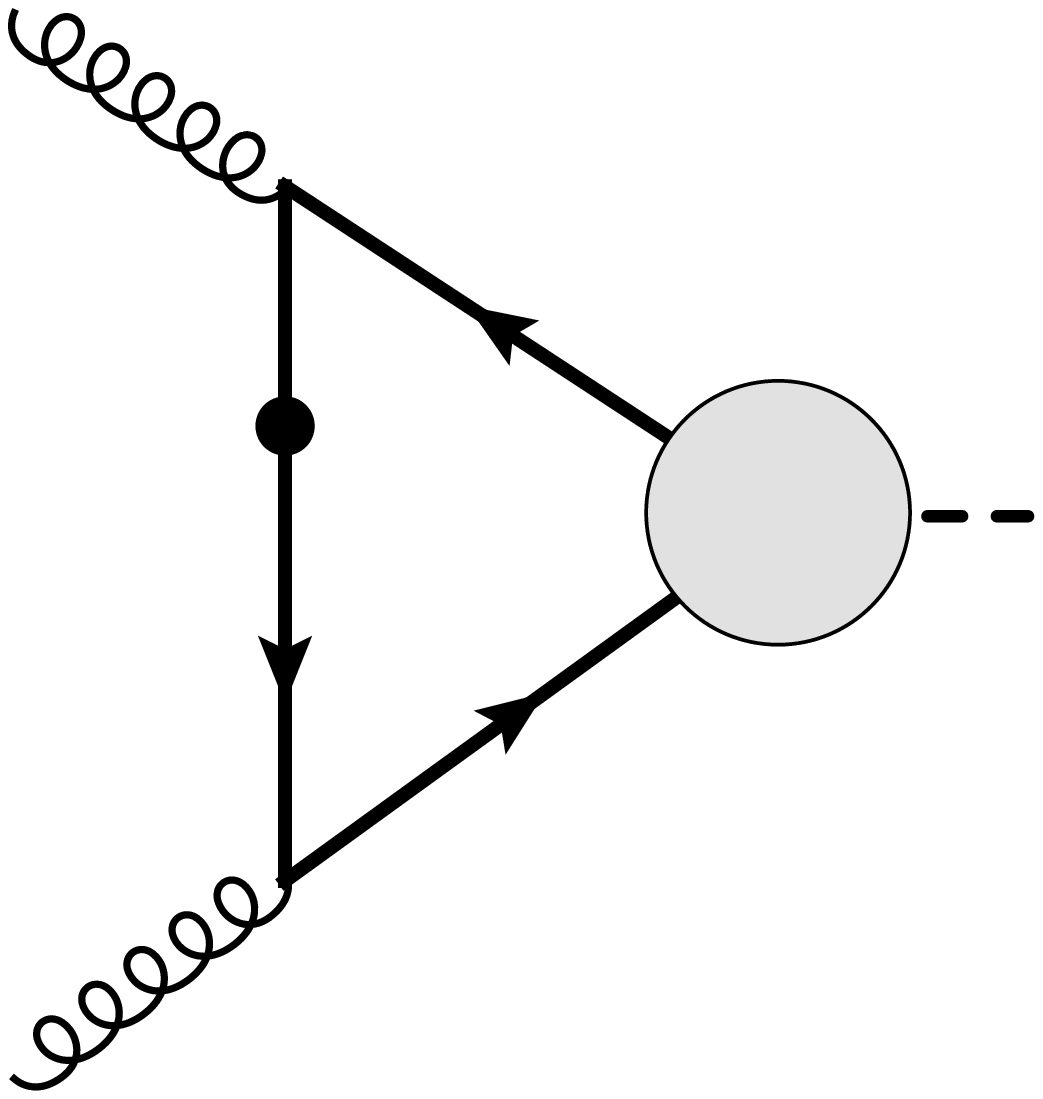}
\hspace*{-9.5mm}\raisebox{12.5mm}{\small ${ m_q^0}$}&
\hspace*{03mm}\includegraphics[width=2.4cm]{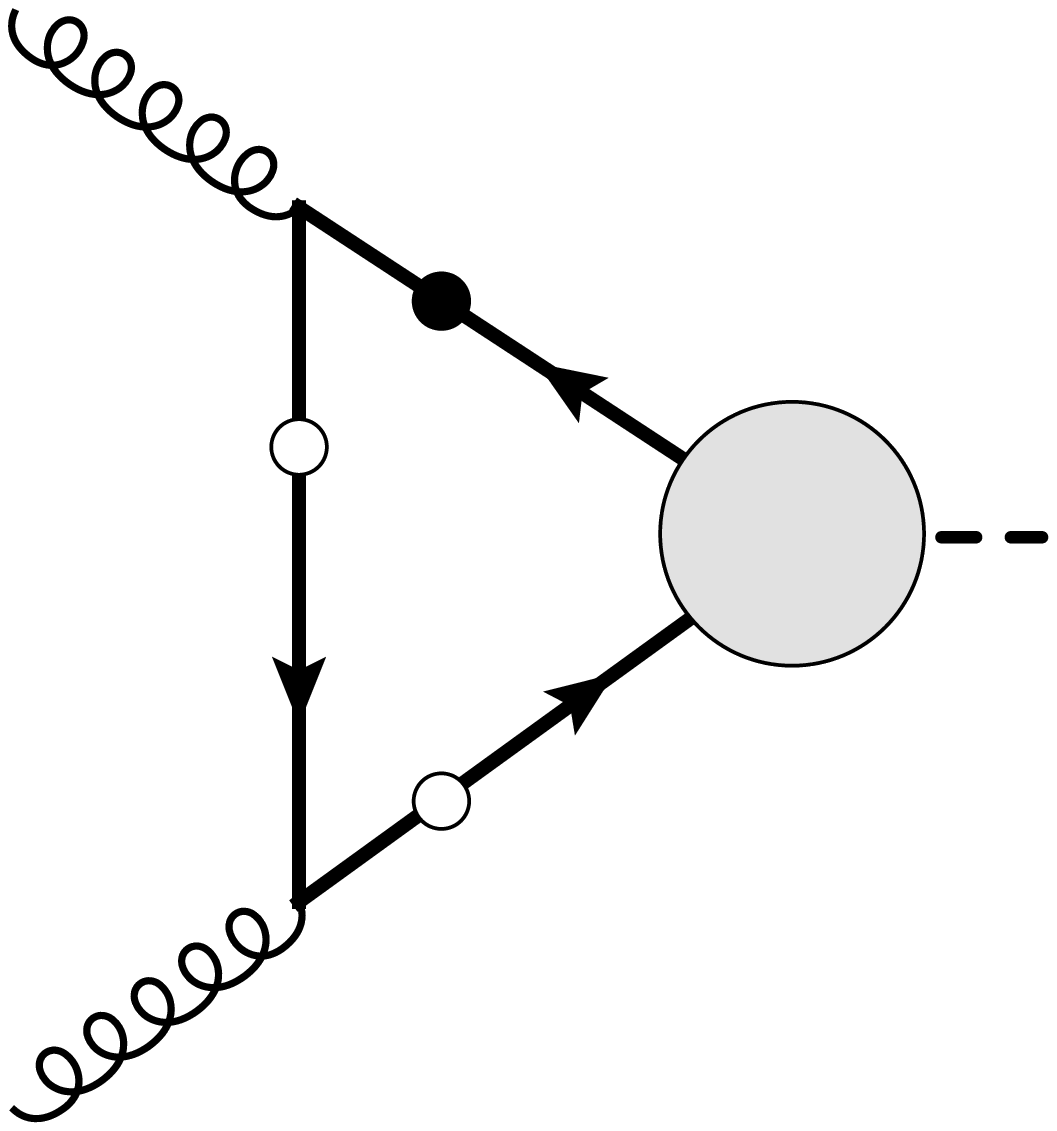}
\hspace*{-9.5mm}\raisebox{12.5mm}{\small ${ m_q^0}$}&
\hspace*{03mm}\includegraphics[width=2.4cm]{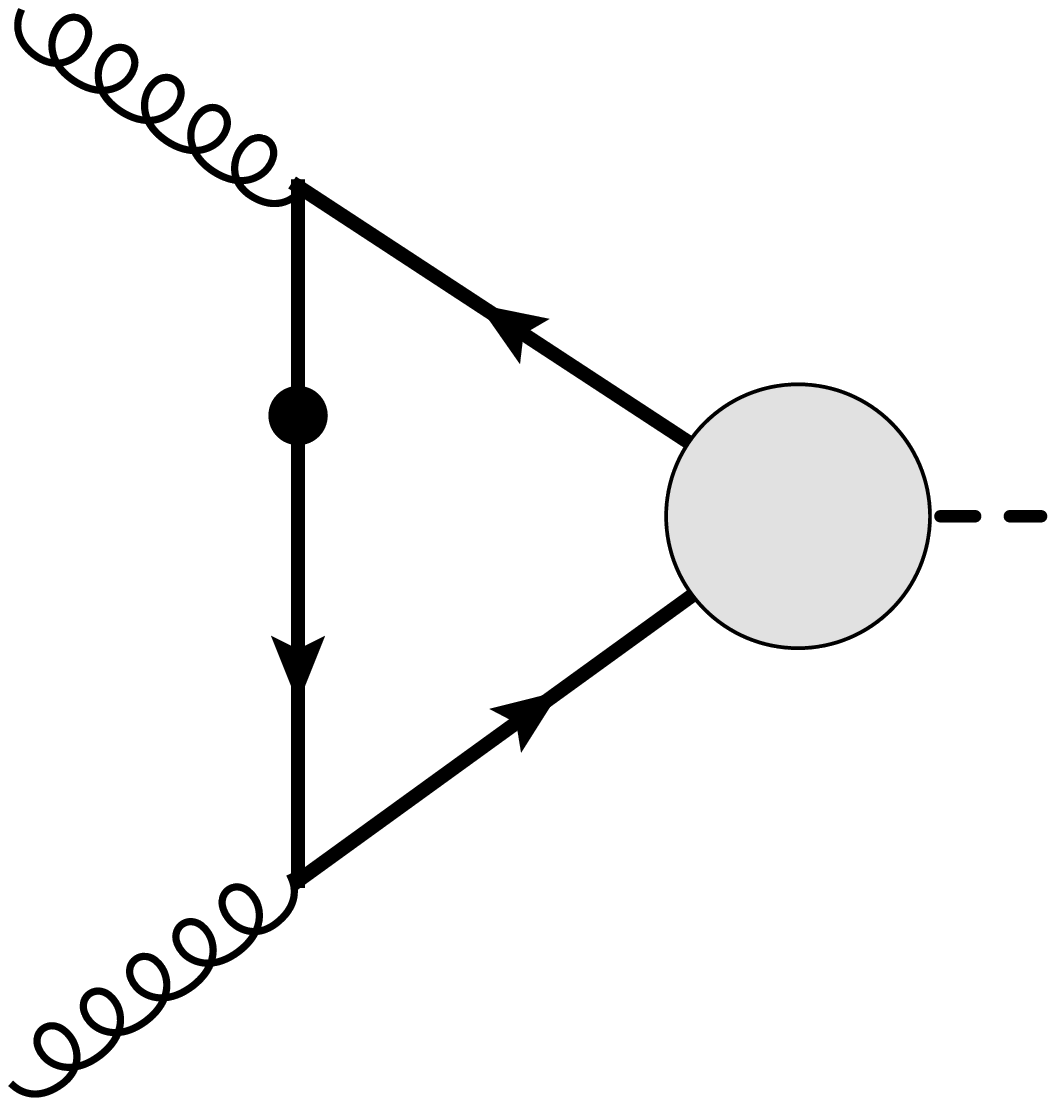}
\hspace*{-9.5mm}\raisebox{12.5mm}{\small ${m_q^2}$}\\
(a)&\hspace*{04mm}(b)&\hspace*{05mm}(c)& \hspace*{05mm} (d) \\
\end{tabular}
\end{center}
\caption{\label{fig::1}  (a)  the leading-order one-loop
Feynman diagram for the ${\cal O}(m_q)$  logarithmic
contribution to the quark  Pauli form factor. The Feynman
diagrams representing  (b) ${\cal O}(m_q)$ and (c), (d)  ${\cal
O}(m_q^3)$  double-logarithmic contributions to the $ggH$
amplitude.  The black (empty) circle represents the mass
(loop momentum) insertion. The gray blobs correspond to the
double-logarithmic off-shell scalar form factor at ${\cal
O}(m_q^{0})$ and ${\cal O}(m_q^{2})$. Symmetric diagrams and
the diagrams with the opposite  direction of the closed quark
line are not shown.}
\end{figure}

The leading-order contribution to the  Pauli form factor is
given by the one-loop vertex diagram and can  be written as follows
\begin{equation}
\left[F_2\right]_{1-loop}={C_F\alpha_s\over \pi}
{1\over 1+4\rho}I_1\,,
\label{eq::F2oneloop}
\end{equation}
where the scalar integral over the virtual gluon momentum
\begin{equation}
I_1={-i}\int{{{\rm d}^4l}\over \pi^2}{(p_1l)+(p_2l)\over
l^2 \left((p_1-l)^2-m_q^2\right)\left((p_2-l)^2-m_q^2\right)}\,
\label{eq::I1def}
\end{equation}
corresponds to a single insertion of the loop momentum in the
numerator of a  quark propagator, Fig.~\ref{fig::1}(a). At
the same time the terms without the loop momentum do not
provide the relevant Lorentz structure. The logarithmically
enhanced corrections to the on-shell (or almost on-shell)
amplitudes in the high-energy limit are universally associated
with the emission of the virtual particles which are soft
and/or collinear to the large external momenta.  For the
one-loop Pauli form factor the virtual gluon momentum in the
numerator cancels one of the eikonal propagators and makes the
integrand not sufficiently singular to develop the
double-logarithmic contribution in the leading order in $\rho$.
Hence the integral generates only   a single soft logarithm
\begin{equation}
I_1=-\ln\rho+\ldots\,,
\label{eq::I1result}
\end{equation}
where the ellipsis stands for the power-suppressed terms. Since
the higher-order Sudakov corrections factor out we get
$F_2^{(1)}=0$  in the double-logarithmic approximation and
\begin{equation}
F_2^{(1)}=-{C_F\alpha_s\over\pi}\ln\rho
\label{eq::F21nll}
\end{equation}
in the next-to-leading logarithmic approximation to all orders
of perturbative expansion. The absence of the
double-logarithmic contribution in the leading-power Pauli form
factor can be easily understood. Indeed, in a physical gauge
the collinear logarithms are generated by the self-energy
corrections  to the on-shell external lines
\cite{Frenkel:1976bj} while a new Lorentz structure in the
quark coupling to the external electromagnetic field  can only
result from a vertex correction. We, however are interested in
the ${\cal O}(\rho^2)$ double logarithms  contributing to
$F_2^{(2)}$. The analysis of such terms is more complicated
since the formal expansion of the integrand in
Eq.~(\ref{eq::I1def}) results in more singular integrals, which
may have double-logarithmic scaling. A systematic way to study
the mass-suppressed double logarithms has been suggested in
\cite{Penin:2014msa} and discussed in detail in
\cite{Penin:2016wiw}. It is based on the expansion by regions
approach \cite{Beneke:1997zp,Smirnov:1997gx,Smirnov:2002pj}
which gives the coefficients of the small-mass expansion in
terms of the singular homogeneous integrals. The coefficient of
the double-logarithmic term can be read of the highest
singularity of these integrals. This singularity in turn can be
obtained by the classical method of Sudakov
\cite{Sudakov:1954sw} which on its own is  blind to the power
corrections. As it has been shown in \cite{Penin:2014msa} the
exchange by {\it massless} soft gauge boson does not produce
double logarithms in the first order in $\rho$. It can be
directly checked for the one-loop integral which gives
\begin{equation}
\left[F_2^{(2)}\right]_{1-loop}={\cal O(\ln\rho)}\,,
\label{eq::F21oneloop}
\end{equation}
and the result extends to an arbitrary number of the soft gluon
exchanges (see \cite{Penin:2016wiw} for detailed discussion).
We would like to emphasize that the above  statement is valid
only for the massless soft particles. A presence of a mass in
the soft propagator does lead to the double-logarithmic
corrections at ${\cal O(\rho)}$ as we will see in the next
section.

\begin{figure}[t]
\begin{center}
\begin{tabular}{cccc}
\includegraphics[width=2.4cm]{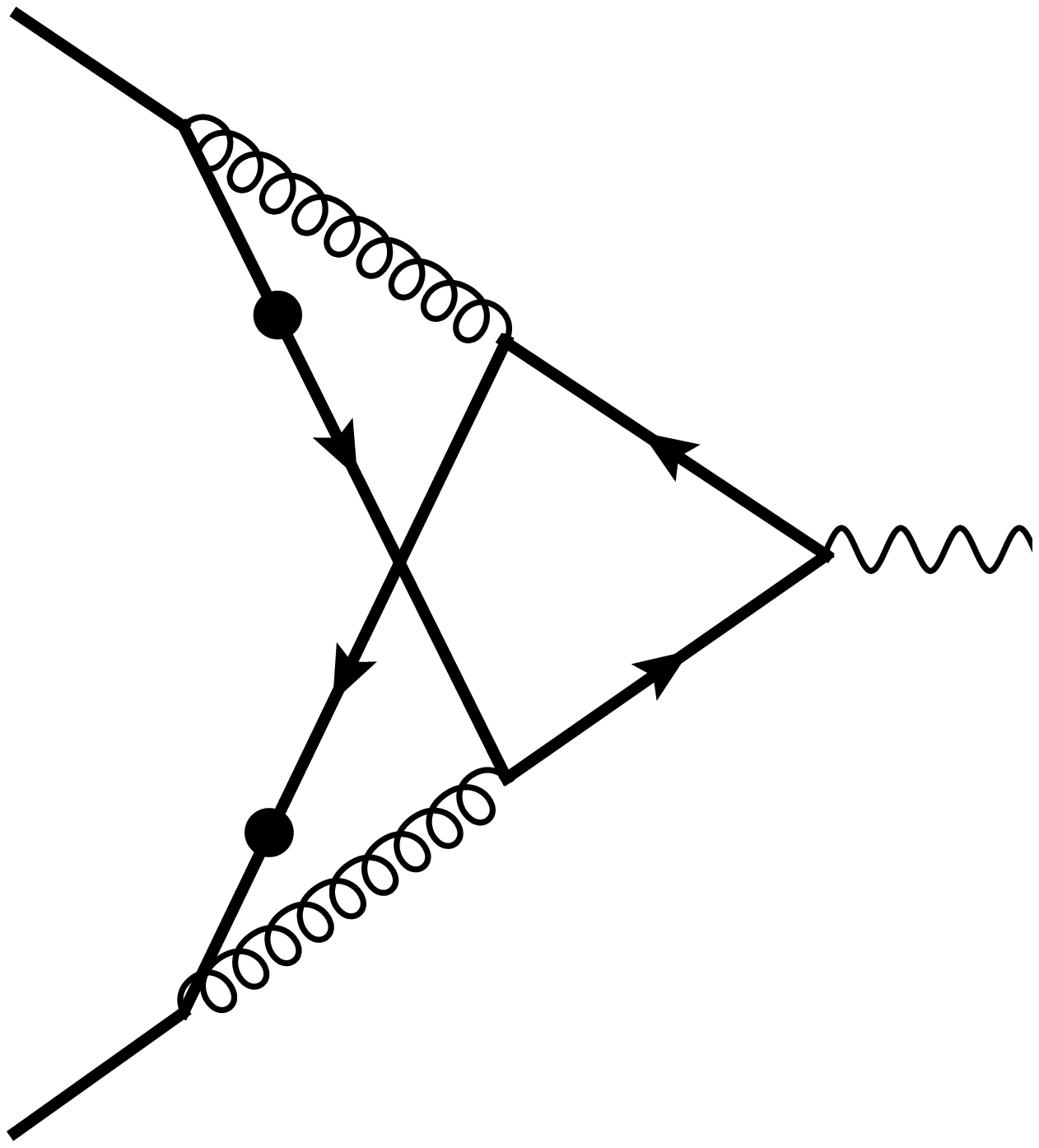}&
\hspace*{6mm}\includegraphics[width=2.4cm]{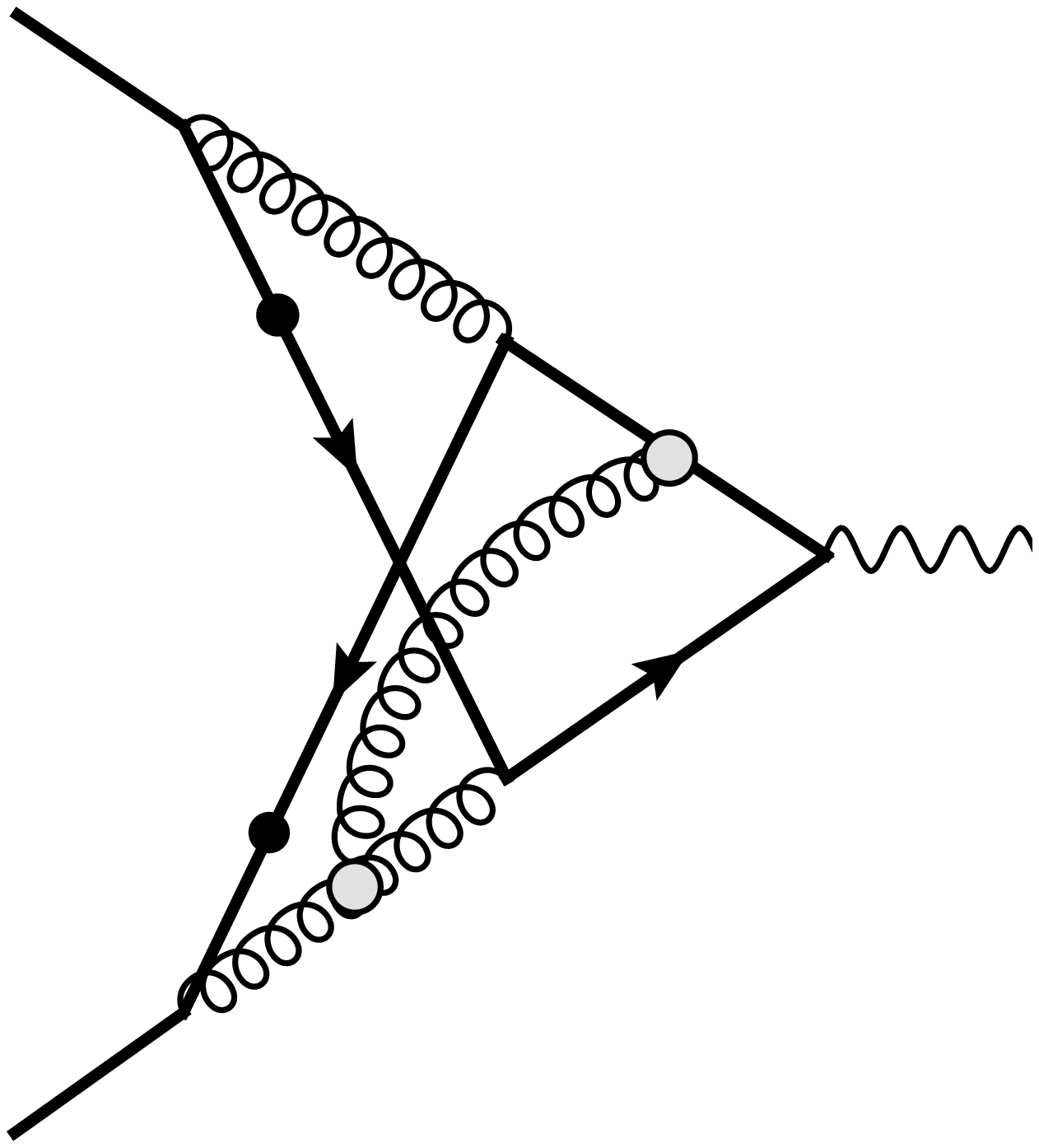}&
\hspace*{6mm}\includegraphics[width=2.3cm]{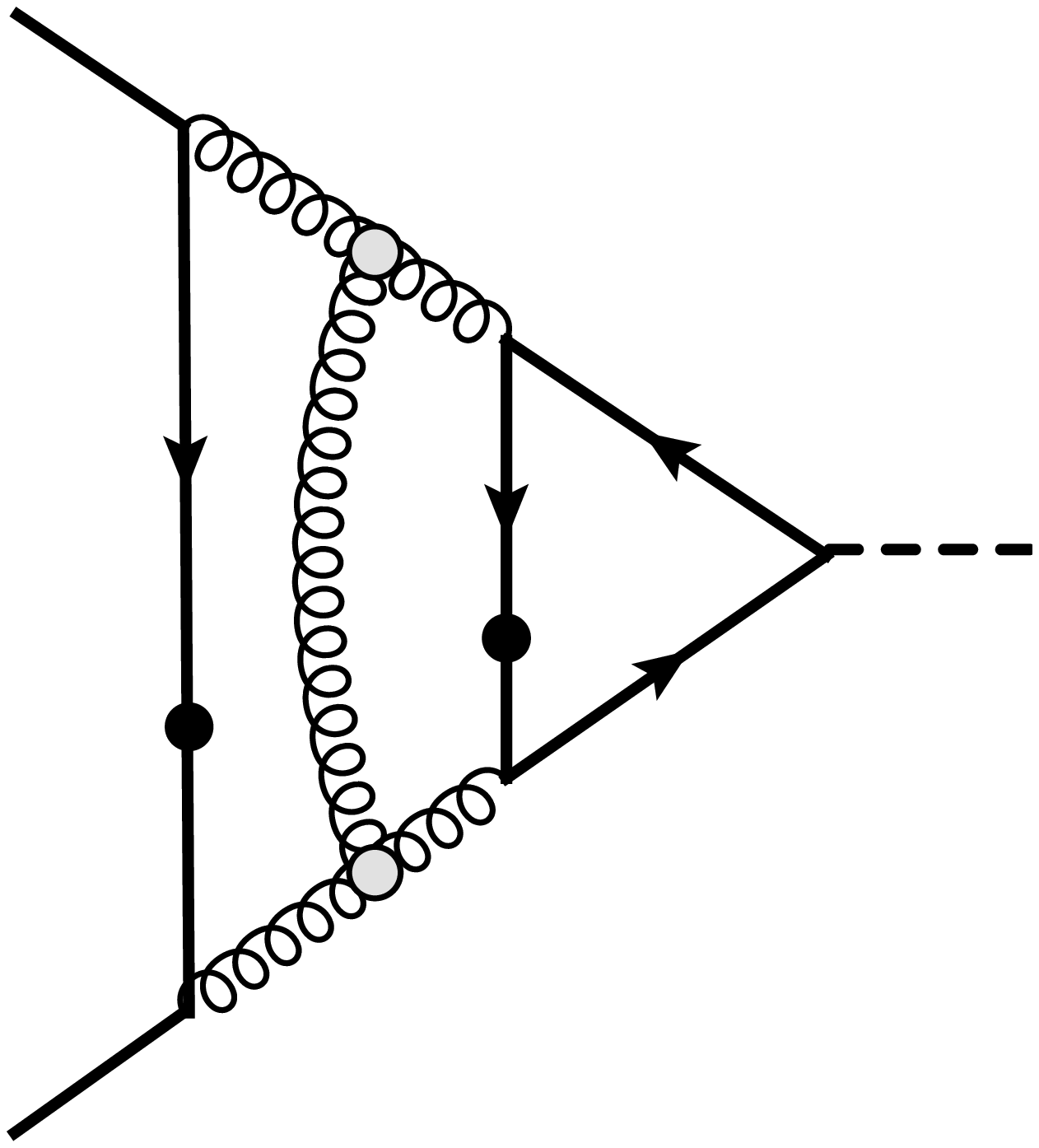}&
\hspace*{6mm}\includegraphics[width=2.3cm]{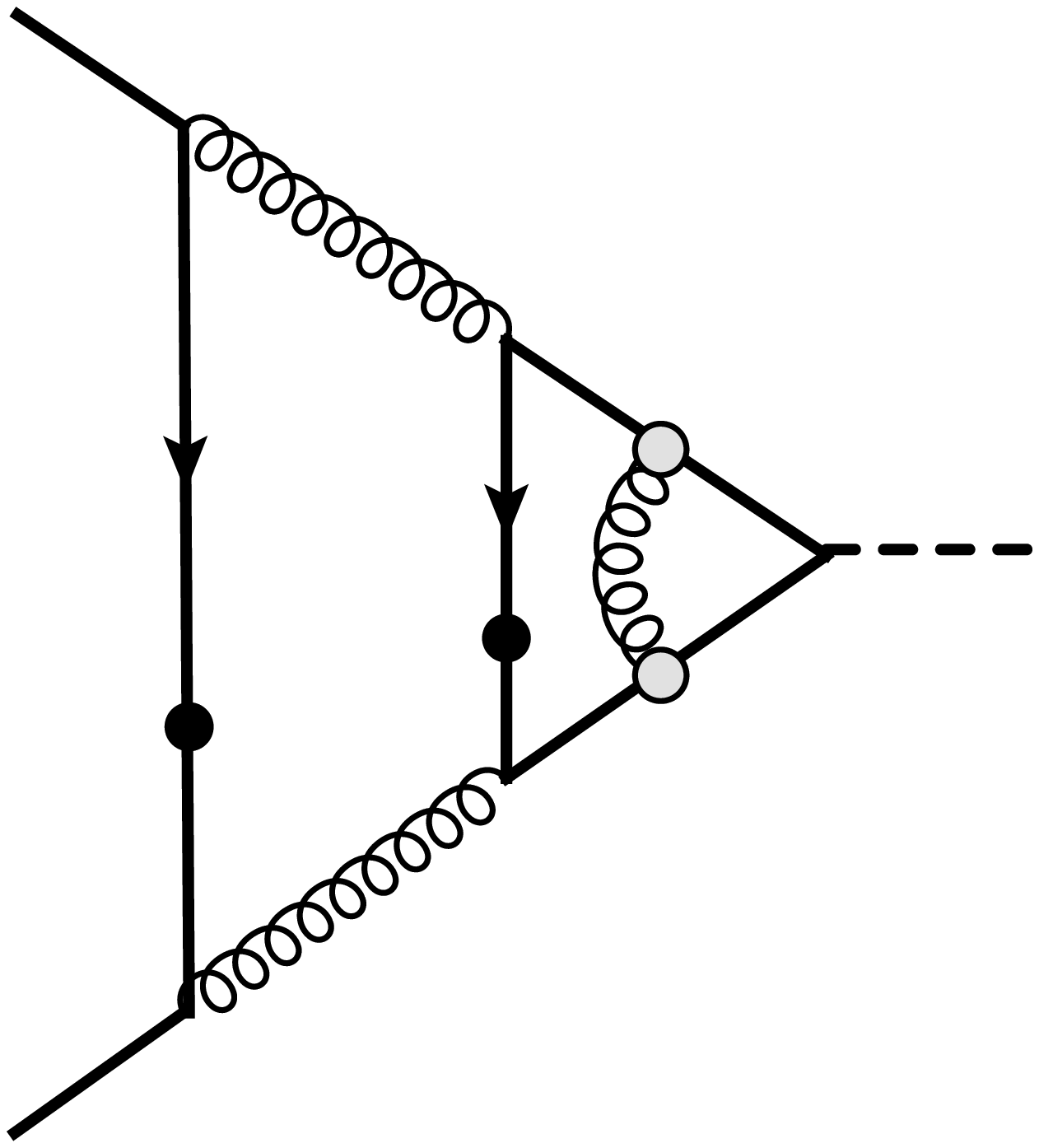}\\
(a)&\hspace*{6mm}(b)&\hspace*{6mm}(c)&\hspace*{6mm}(d)\\
\end{tabular}
\end{center}
\caption{\label{fig::2}  (a) the leading two-loop Feynman
diagram for the ${\cal O}(m_q^3)$ double-logarithmic correction
to the quark Pauli form factor. The diagrams with an effective
soft gluon exchange which incorporate the non-Sudakov
double-logarithmic corrections (b) to the  ${\cal O}(m_q^3)$
Pauli form factor and (c), (d) to the ${\cal O}(m_q^2)$ scalar
form factor of a quark. The effective vertices (gray circles)
are defined in the text.}
\end{figure}

Thus the soft gluon exchanges do not contribute to $F_2^{(2)}$
in the double-logarithmic approximation. At the same time as in
the case of the Dirac form factor  at ${\cal O}(m_q^{2})$
\cite{Penin:2014msa} starting with two loops the
double-logarithmic contribution to this coefficient is
generated through the soft virtual quark pair exchange,
Fig.~\ref{fig::2}(a), where the mass suppression factor comes
from the numerators of the soft propagators, {\it i.e.} is
associated with the chirality flip. This makes the soft quark
propagators sufficiently singular to produce the
double-logarithmic contribution. In Fig.~\ref{fig::2}(a)  the
large external momenta flow through the edges of the diagram
which for the soft loop momenta form the eikonal lines. Thus
the corresponding momentum configuration is the opposite of the
standard Sudakov case with soft gauge bosons and eikonal
fermions. The  Pauli form factor structure requires an
additional chirality flip on an eikonal quark line, which is
provided either by the mass term of the propagator or by an
external momentum $\slashed{p}_i$ acting on the corresponding
on-shell quark field. Such terms cancel in the  diagrams with
the soft gluon exchange but give a contribution which is not
suppressed  by the soft momentum in the diagram
Fig.~\ref{fig::2}(a),  where  due to the topology of the
fermion flow the operators $\slashed{p}_i$ should be commuted
with the photon vertex before the equation of motion for the
initial and final quark states can be applied. Let us consider
the origin of the relevant Lorentz structure in more detail.
Due to chirality conservation the eikonal gauge bosons must
have transversal polarization. By using the property
$\gamma_i\gamma_j\gamma_i\gamma_j=0$ of the Dirac matrices in
two-dimensional transversal space we find that after neglecting
the virtual momenta the entire contribution to the Pauli form
factor is generated  by  the photon vertex
$\slashed{p}_2\gamma_\mu \slashed{p}_1$ part of the Dirac chain
where one of the external momenta is converted into $m_q$ by
the equation of motion. The evaluation of the corresponding
two-loop double-logarithmic integral is discussed in detail in
\cite{Penin:2014msa,Penin:2016wiw} and gives
\begin{equation}
\left[F_2^{(2)}\right]_{2-loop}={2\over 3}{C_F(C_A-2C_F)}x^2+\ldots\,,
\label{eq::F21twoloop}
\end{equation}
where the ellipsis stands for the subleading logarithms, which
agrees with the expansion of the exact result
\cite{Bernreuther:2004ih}. The higher-order double-logarithmic
corrections are generated by the multiple  exchanges of the
leading-power soft gluons with light-cone
polarization.\footnote{A covariant gauge is implied.}  The
emission of a soft gluon with polarization $\alpha$ off the
eikonal quark is trivial in the spinor space due to the
identity
\begin{equation}
\left(\slashed{p}_i+m_q\right)\gamma^\alpha
\left(\slashed{p}_i+m_q\right)
= 2p_i^\alpha\left(\slashed{p}_i+m_q\right).
\label{eq::softvertex}
\end{equation}

\noindent
Thus the higher-order double-logarithmic corrections do not
affect the  spinor part of the amplitude  and the Lorentz
structure relevant for the Pauli form factor is generated   by
the same vertex part of the Dirac chain  as in two loops. As a
consequence they are identical to the double-logarithmic
corrections to the coefficient $F_1^{(1)}$  stemming from  the
same diagram Fig.~\ref{fig::2}(a). After factoring out the
Sudakov corrections to the external quark lines, the remaining
non-Sudakov double logarithms are described by the diagram
Fig.~\ref{fig::2}(b) \cite{Liu:2017vkm,Liu:2018czl}. The
effective soft gluon exchange in this diagram exponentiate and
its coupling is obtained from the standard QCD expression  by
replacing the quadratic Casimir operator $C_A$ of the adjoint
representation (gluon color charge) with the difference
$C_A-C_F$ which reflects the nonconservation of the color
charge along the eikonal lines in the process with soft quark
exchange, which is the physical origin of the non-Sudakov
corrections. The result for the $F_2^{(2)}$  then reads
\begin{equation}
F_2^{(2)}={2\over 3}{C_F(C_A-2C_F)}x^2f\left(-z\right)
\label{eq::F2result}
\end{equation}
with the function $f(z)$ given by Eq.~(\ref{eq::f}). The Taylor
expansion of this function
\begin{equation}
f(z)=1+{z\over 5}+{11\over 420}z^2+{z^3\over 378}+\ldots
\label{eq::fataylor}
\end{equation}
gives the loop-by-loop double-logarithmic approximation of the
Pauli form factor.  In the limit of large momentum transfer the
asymptotic behavior of the form factor crucially depends on the
gauge group. The variable $z$ is negative in QED and positive
in QCD. The relevant asymptotic expressions at  $z\to\infty$
are respectively
\begin{equation}
f(z)\sim 6\left[\ln\left({z\over 2}\right)+\gamma_E\right]
\left({2\pi e^{z}\over z^5}\right)^{1/2}\!\!
\label{eq::fasymQED}
\end{equation}
and
\begin{equation}
f(-z)\sim   \left[\left(\ln{(2z)}+\gamma_E\right)^2
-{\pi^2\over 2}\right]{3 \over z^2}\,,
\label{eq::fasymQCD}
\end{equation}
where $\gamma_E=0.577215\ldots$ is the Euler constant. The
details of the derivation  of the above asymptotic formulae are
given in Appendix~\ref{sec::appC}.\footnote{In
\cite{Liu:2017vkm} the coefficients of the leading powers of
logarithms in Eqs.~(\ref{eq::fasymQED},\ref{eq::fasymQCD}) have
been estimated numerically with a rather low accuracy of the
fit.} Thus in
QED the Pauli form factor at ${\cal O}(m_q^{3})$ has the
leading  asymptotic behavior given by the exponential factor
$e^{x/2}$. At the same time in QCD it scales with the
double-logarithmic variable as $\ln^2 x$.

\section{Higgs boson production in gluon fusion}
\label{sec::3}
A quark loop mediated $ggH$ amplitude can be written
as follows
\begin{equation}
{\cal M}^q_{gg H}=T_F{\alpha_s\over \pi}
{y_qm_q\over m_H^2}\left(p_1^\mu p_2^\nu- g^{\mu\nu}(p_1p_2)\right)
A^a_\nu(p_1)A^a_\mu(p_2){H}{M}^q_{gg H}\,,
\label{eq::MggHdef}
\end{equation}
where $y_q$ is the quark Yukawa coupling, $m_H$ is the Higgs
boson mass, $p_i^2=0$, $(p_1p_2)=-m_H^2/2$, the  gauge
condition $\partial^\mu A^a_\mu=0$ is implied and one can choose
the transversal polarization of the gluon fields.  In the heavy
quark limit $m_q\gg m_H$ the scalar amplitude approaches the
value ${M}^q_{\gamma\gamma H}= -2/(3\rho)$, where now $\rho=
m_q^2/m_H^2$ is  a Minkowskian parameter. In the opposite limit
of light quark $m_q\ll m_H$ it can be expanded in an asymptotic
series
\begin{equation}
{M}^q_{ggH}= Z_g^2\sum_{n=0}^\infty \rho^n
M^{(n)}_{ggH} \,,
\label{eq::MggH}
\end{equation}
where the coefficients $M^{(n)}_{ggH}$ are finite and
\begin{equation}
Z_{g}^2=\exp\left[{-{C_As^{-\varepsilon}\over\varepsilon^2}
{\alpha_s\over 2\pi}}\right]
\label{eq::Zg}
\end{equation}
is  the universal Sudakov factor for the  external on-shell
gluon  lines which incorporates all the infrared divergencies
of the amplitude. Note that as in the case of Pauli form factor
the amplitude is loop generated and in the high-energy
(small-mass) limit is suppressed by the quark mass due  to
chirality flip at the Higgs boson vertex.

The leading-order one-loop scalar amplitude reduces to
\begin{equation}
\left[{M}^q_{ggH}\right]_{1-loop}=2J_1+8J_2\,,
\label{eq::MHoneloop}
\end{equation}
where
\begin{equation}
J_1={i }\int{{\rm d}^4l\over \pi^2} {2(p_1p_2)\over (l^2-m_q^2)
\left((p_1-l)^2-m_q^2\right)\left((p_2-l)^2-m_q^2\right)}\,
\label{eq::J1def}
\end{equation}
and
\begin{equation}
J_2= i\int{{\rm d}^4l \over \pi^2 }{ l^2-4(lp_1)(lp_2)/(p_1p_2)
\over (l^2-m_q^2) \left((p_1-l)^2-m_q^2\right)
\left((p_2 -l)^2-m_q^2\right)}\,
\label{eq::J2def}
\end{equation}
are  the scalar integrals corresponding to no and double
insertion of the loop (soft quark) momentum $\slashed{l}$ in
the numerators of the quark propagators, respectively ({\it
cf.} the diagrams in Fig.~\ref{fig::1}(b) and
Fig.~\ref{fig::1}(c) with the leading-order Higgs boson
vertex). The integral $J_1$  is responsible for the leading
double-logarithmic contribution to the ${\cal O}(m_q)$
coefficient $M^{(0)}_{ggH}$. For the further analysis it is
instructive to recall the evaluation of this contribution (see
{\it e.g.} \cite{Liu:2018czl}). With the double-logarithmic
accuracy  the propagators of the soft and eikonal quarks can be
approximated as follows
\begin{eqnarray}
&& {1\over l^2-m_q^2}  \approx
- i \pi  \delta(l^2-m_q^2)\,,
\label{eq::propsoft}\\
&&{1\over (p_i-l)^2-m_q^2}\approx
-\frac{1}{2(p_il)}\,.
\label{eq::propeik}
\end{eqnarray}
Then the standard Sudakov parametrization  of the  soft quark
momentum $l=up_1+vp_2+l_\perp$ is introduced, where the first
two terms correspond to the light-cone components and the last
term corresponds to the transversal component in the plane
orthogonal to the gluon momenta. The validity of the eikonal
approximation in Eq.~(\ref{eq::propeik}) requires $|u|,|v|<1$
and the additional kinematical constraints $|uv|>\rho$  has to
be imposed to ensure that the soft quark propagator in
Eq.~(\ref{eq::propsoft}) can go on the mass shell. After
integrating Eq.~(\ref{eq::J1def}) over  ${l}_\perp$ with the
double-logarithmic accuracy we get
\begin{equation}
J_1\approx \int_{\rho}^{1}{{\rm d}v\over v}
\int_{\rho/v}^{1}{{\rm d}u\over u}=\ln^2\!\rho\int_0^1
{\rm d}\xi \int_{0}^{1-\xi}{\rm d}\eta
={\ln^2\!\rho\over 2}\,,
\label{eq::J1uv}
\end{equation}
where the normalized logarithmic variables  $\eta=\ln v/\ln\rho
$ and $\xi=\ln u/\ln\rho$ are introduced.\footnote{The
contributions of the positive and negative Sudakov parameters
are symmetric so we rewrite the total integral in terms of
positive $u$ and $v$.}  As in the case of the soft  gluon
exchange the expansion of $J_1$ to ${\cal O}(\rho)$ does not
result in a double-logarithmic contribution. However, in the given
kinematics this can be seen immediately since $p_i^2=0$. Indeed,
the expansion of the eikonal quark propagators then reads
\begin{equation}
{1\over (p_i-l)^2-m_q^2}=
-\frac{1}{2(p_il)}\left(1+\frac{l^2-m_q^2}{2(p_il)}
+\ldots\right)
\label{eq::propexp}
\end{equation}
so that all the subleading  terms cancel soft quark propagator
and have no double-logarithmic scaling. Thus we get
\begin{equation}
J_1={1\over 2}\left(\ln^2\!\rho+{\cal O}(\rho \ln\rho)\right)+\ldots\,,
\label{eq::J1res}
\end{equation}
which in double-logarithmic approximation gives
\begin{equation}
\left[M^{(0)}_{ggH}\right]_{1-loop}=\ln^2\!\rho
\label{eq::M0oneloop}
\end{equation}
and does not contribute to $M^{(1)}_{ggH}$.

The case of the integral Eq.~(\ref{eq::J2def}) is less trivial.
It remains finite at $m_q\to 0$ but its expansion to  ${\cal
O}(\rho)$ does produce a double-logarithmic contribution. The
second term in the numerator of Eq.~(\ref{eq::J2def}) cancels
both eikonal quark propagators up to the terms proportional to
$l^2-m_q^2$ and can be omitted. In the first term for the
on-shell soft quark we can replace $l^2$ with $m_q^2$ and get
the same double-logarithmic integral as for $J_1$ up to an
overall factor $-\rho$, which gives
\begin{equation}
J_2=-{1\over 2}\left( 1+  \rho \ln^2\!\rho+\ldots\right)\,,
\label{eq::J2res}
\end{equation}
and
\begin{equation}
\left[M^{(1)}_{ggH}\right]_{1-loop}=-4\ln^2\!\rho\,.
\label{eq::M1oneloop}
\end{equation}

\begin{figure}[t]
\begin{center}
\begin{tabular}{cccc}
\includegraphics[width=3.2cm]{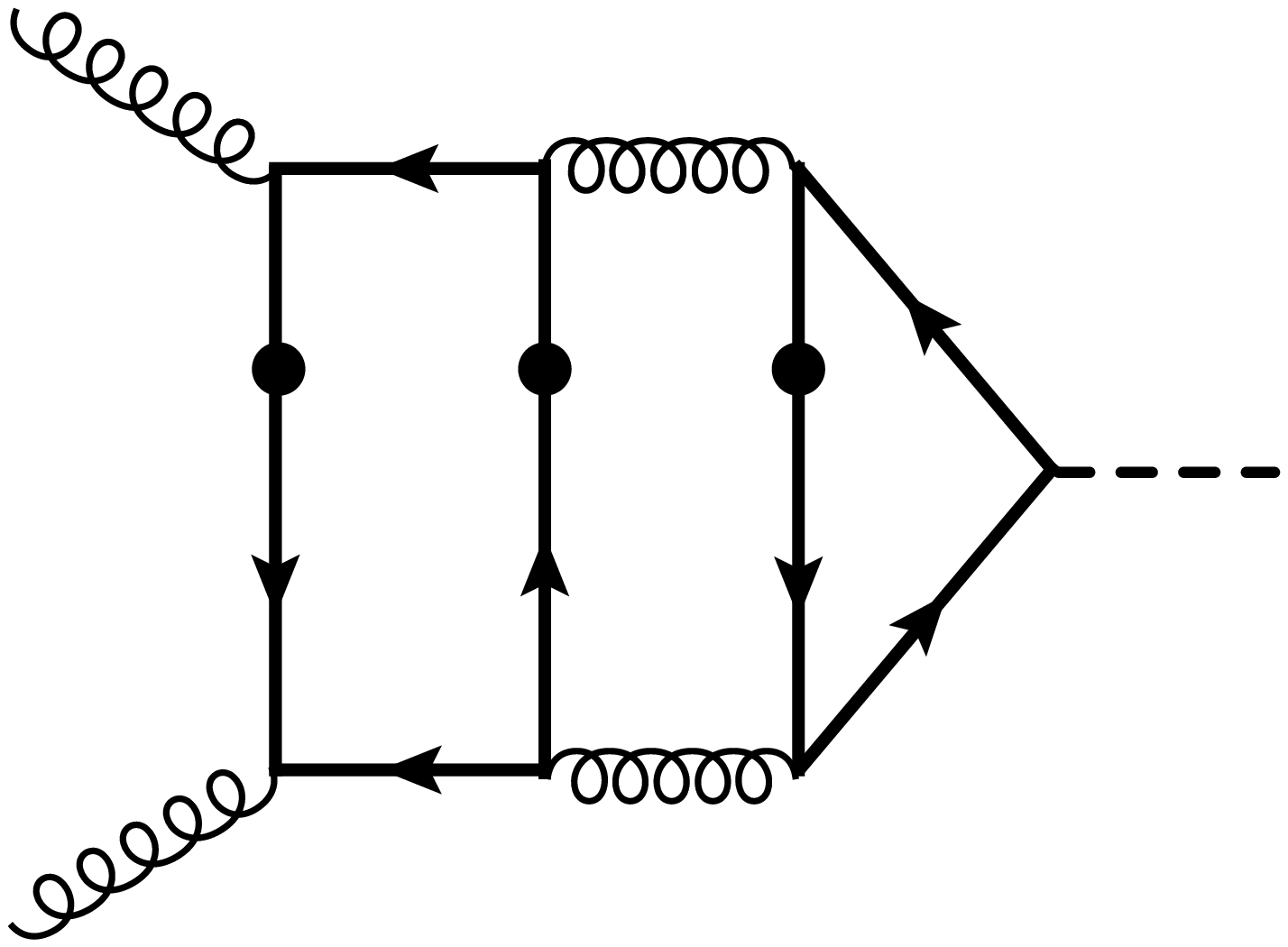}&
\hspace*{2mm}\includegraphics[width=3.2cm]{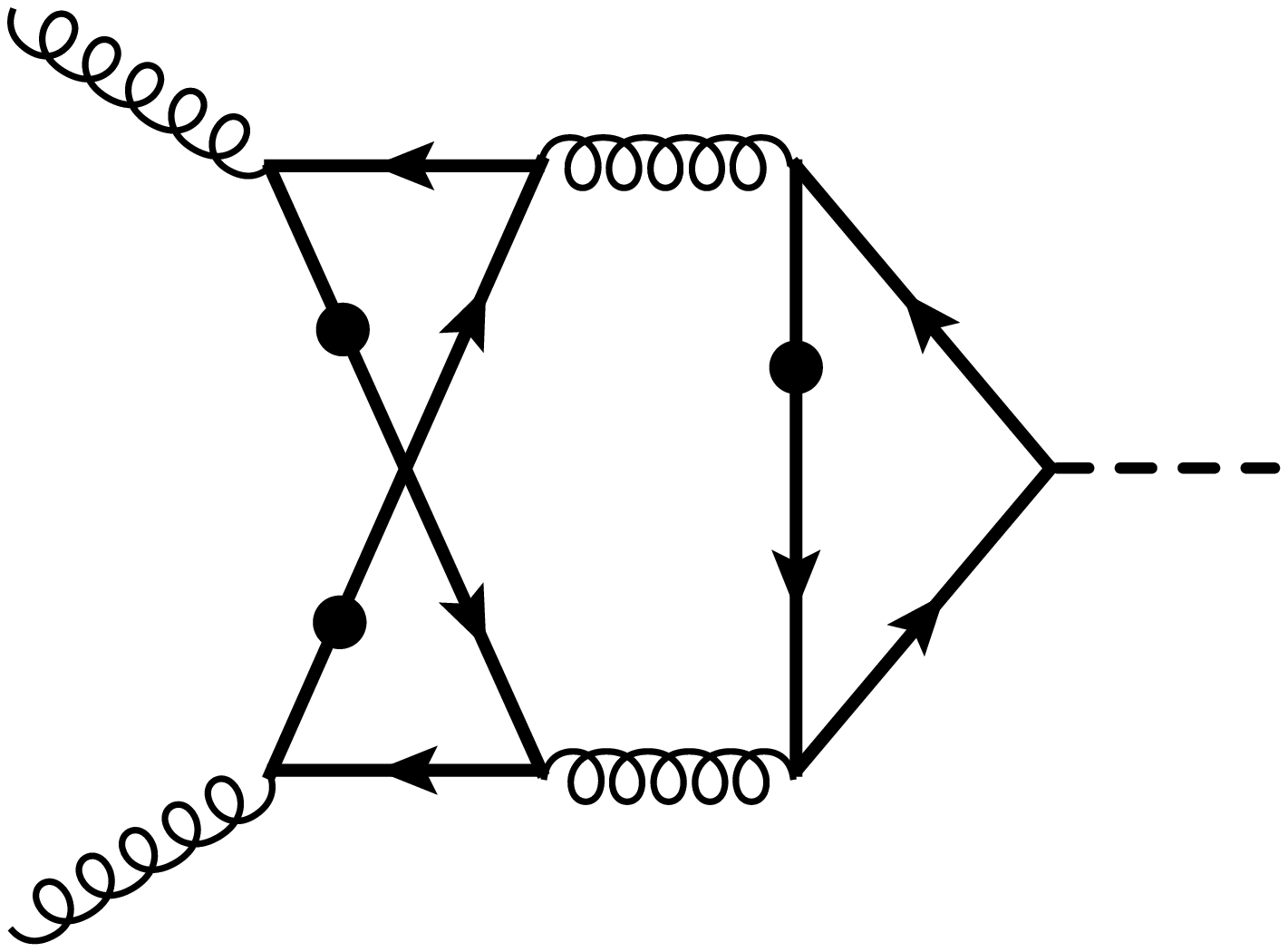}&
\hspace*{2mm}\includegraphics[width=3.2cm]{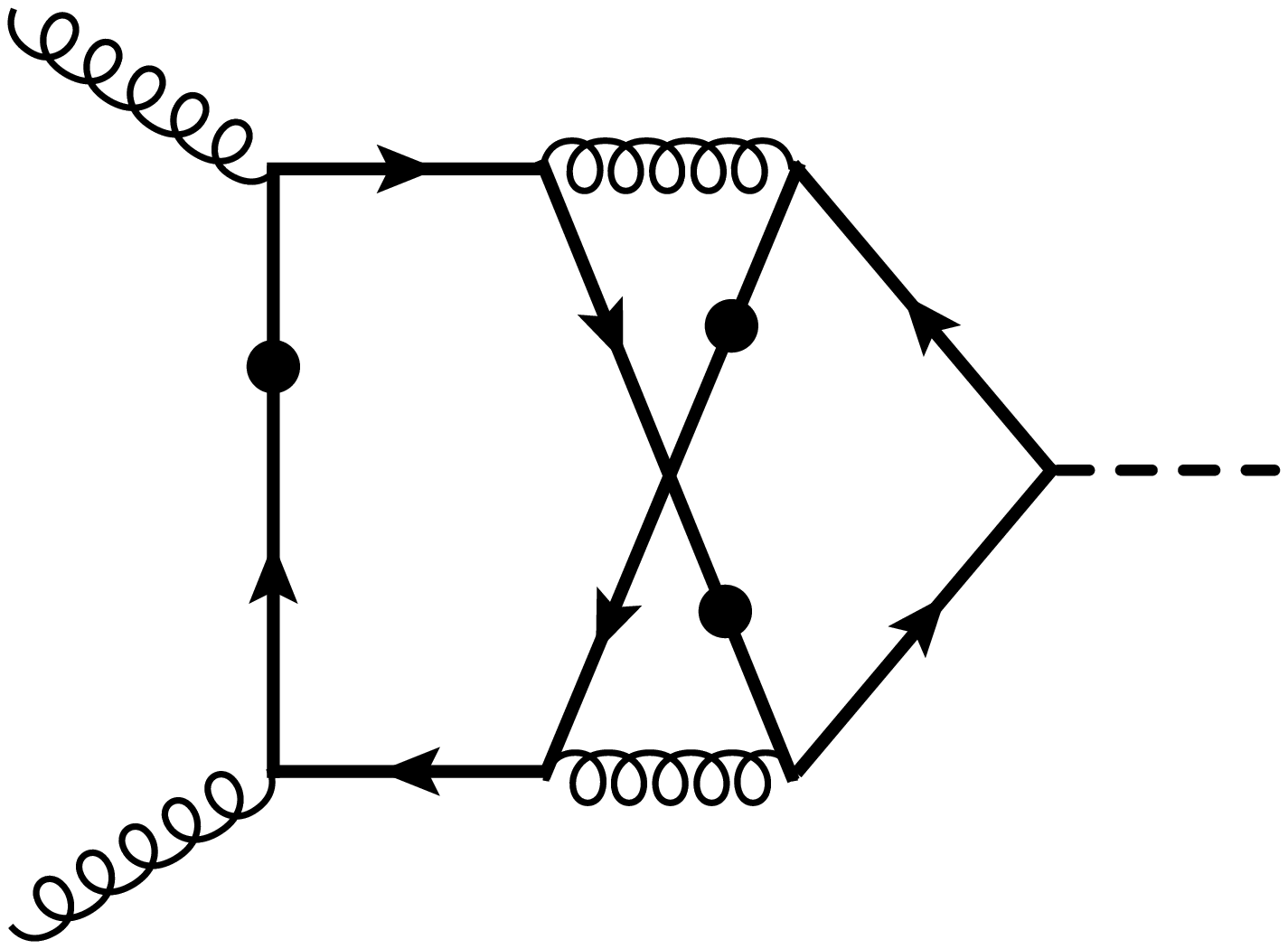}&
\hspace*{2mm}\includegraphics[width=3.2cm]{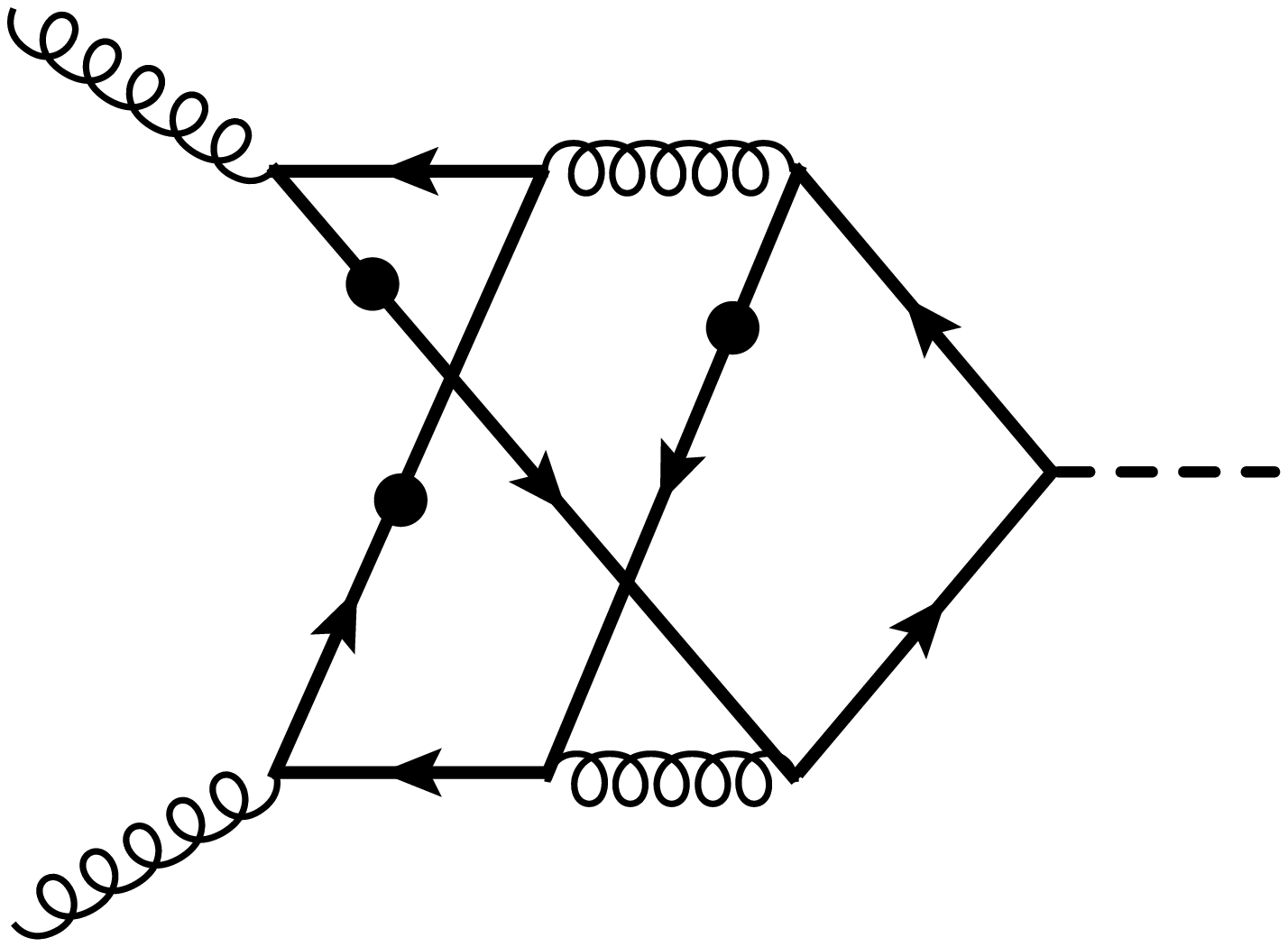}\\
(a)&\hspace*{2mm}(b)&\hspace*{2mm}(c)&\hspace*{2mm}(d)\\
\end{tabular}
\end{center}
\caption{\label{fig::3}  The three-loop Feynman diagrams for the
Higgs boson two-photon decay amplitude with triple  soft quark
exchange.}
\end{figure}

\noindent
Let us consider the higher-order double-logarithmic
contributions due to dressing of the one-loop diagram with
multiple leading-power gluon exchanges. After factoring out the
external line Sudakov corrections into $Z_g^2$ the remaining
non-Sudakov mass logarithms are  determined by the effective
soft gluon corrections to the Higgs boson vertex which  can be
obtained from the double-logarithmic result  for  the off-shell
quark form factor given in the Appendix~\ref{sec::appA}. At
${\cal O}(m_q)$ only the leading-power term should be kept in
Eq.~(\ref{eq::FSseries}) and  included into the one-loop
integrand of Eq.~(\ref{eq::J1uv}), see Fig.~\ref{fig::1}(b). To
account for the variation of the color charge along the eikonal
lines in the process of soft quark emission the color weight in
the Sudakov factor Eq.~(\ref{eq::Zqetaxi}) for the off-shell
quark lines should be changed from $C_F$ to $C_F-C_A$
\cite{Liu:2017vkm,Liu:2018czl}. This gives
\begin{equation}
M^{(0)}_{ggH}=\ln^2\!\rho\, g(z)\,,
\label{eq::M0}
\end{equation}
where
\begin{equation}
g(z)=2\int_0^1 {\rm d}\xi \int_{0}^{1-\xi}
{\rm d}\eta e^{2z\eta\xi}=
{}_2F_2\left(1,1;{3/2},2;{z/2}\right)\,
\label{eq::g}
\end{equation}
is the  generalized hypergeometric function with the Taylor
expansion
\begin{equation}
g(z)=2\sum_0^\infty {n!\over (2n+2)!}(2z)^n\,.
\label{eq::gseries}
\end{equation}
At ${\cal O}(m_q^3)$ the loop momentum insertion do not affect
the structure of the leading-power  soft gluon contribution
which is represented by the diagram in Fig.~\ref{fig::1}(c).
As it was discussed above, the logarithmic integral over the
soft quark momentum is identical to the ${\cal O}(m_q)$ term
and  the corresponding contribution to $M^{(1)}_{ggH}$ is given
by $-4 M^{(0)}_{ggH}$,  {\it cf.} Eqs.~(\ref{eq::M0oneloop})
and (\ref{eq::M1oneloop}).
In principle in this order of the small-mass expansion one has
to consider the leading power correction to the off-shell form
factor itself. However, as it has been pointed out the soft
gluons do not generate the  double-logarithmic  ${\cal
O}(\rho)$ contribution.  At the same time the power
corrections in the off-shell quark momenta $\Delta_i/m_H^2$ to
Eq.~(\ref{eq::FSseries}) cancel an eikonal quark propagator up
to the terms proportional to $l^2-m_q^2$ and also can be
omitted.

\begin{table}[t]
  \begin{center}
    \begin{tabular}{|c|c|c|c|c|c|c|c|c|}
     \hline
      $n$ & $1$ & $2$ & $3$ & $4$ & $5$ & $6$ & $7$  &$8$  \\
      \hline
      & &  & &  &  &  &  &  \\[-3mm]
      $n^2 2^nn!h_n$ &  ${3\over 7}$ & ${8\over  9}$ & ${90\over 77}$&
      ${59392\over 45045}$ & ${5360\over 3861}$ &
      ${7559936\over 5360355}$ & ${583744\over 415701}$&
      ${2110652416\over 1527701175}$
      \\[1.5mm]
      $n!j^{\rm ab}_n$ &  ${17\over  28}$ & ${83\over 175}$&
      ${241\over 550}$ & ${47984\over 105105}$ &
      ${3645\over 7007}$ & ${97228\over 153153}$&
      ${772588\over 944775}$ & ${19563776\over 17782765}$
      \\[1.5mm]
      \hline
    \end{tabular}
   \end{center}
    \caption{\label{tab::1} The   normalized coefficients of
    the Taylor series for the function $h(z)$ and $j^{\rm
    ab}(z)$ up to  $n=8$.}
\end{table}

Starting with three loops the diagrams with triple  soft quark
exchange, Figs.~\ref{fig::3}(a-d), may contribute to  ${\cal
O}(m_q^3)$ amplitude. The double-logarithmic part of the
diagrams Figs.~\ref{fig::3}(b-d) vanishes after taking the
spinor trace over the closed quark loop. At the same time the
diagram Fig.~\ref{fig::3}(a) includes  a two-loop subdiagram
corresponding to the double-logarithmic  off-shell scalar form
factor  Eq.~(\ref{eq::FSseries}), {\it i.e.} has the structure
of Fig.~\ref{fig::1}(d). To get the corresponding corrections
to the amplitude the next-to-leading power term
$Z_q^2(\eta,\xi)F^{(1)}_S(\eta,\xi)$ should be included  into
the one-loop integrand of  Eq.~(\ref{eq::J1uv}) and as before
in the factor $Z_q^2(\eta,\xi)$ the color weight $C_F$ should
be changed to $C_F-C_A$.  In this way we get the
double-logarithmic corrections to the coefficient
$M^{(1)}_{ggH}$
\begin{equation}
\ln^2\!\rho\,{T_FC_F\over  45}{x^2}h(z)\,,
\label{eq::triple}
\end{equation}
where the function $h(z)$ has the following integral
representation
\begin{equation}
h(z)=6!\int_0^1{\rm d}\eta\int_{0}^{1-\eta}{\rm d}\xi\int_0^\eta{\rm d}
\eta_2\int_{0}^{\xi}{\rm d}\xi_2\int_{0}^{\eta_2}{\rm d}\eta_1\int_{0}^{\xi_2}
{\rm d}\xi_1\,e^{2z\left(\eta\xi- \eta_2\xi_2+ \eta_1\xi_1\right)}\,.
\label{eq::h}
\end{equation}
The coefficients of the Taylor series $h(z)=
1+\sum_{n=1}^\infty h_n z^n$ can be computed for any given $n$
corresponding to the $(n+3)$-loop double-logarithmic
contribution. The  first eight coefficients of the series  are
listed in Table~\ref{tab::1}.

\begin{figure}[t]
\begin{center}
\begin{tabular}{ccc}
\hspace*{-5mm}
\raisebox{24.mm}{\tiny $p_1$}\hspace*{-4.2mm}
\raisebox{3.mm}{\tiny $p_2$}\hspace*{2mm}
\raisebox{8.5mm}{\tiny $l_1$}\hspace*{-3.5mm}
\raisebox{16.mm}{\tiny $l$}\hspace*{-4mm}
\includegraphics[width=2.9cm]{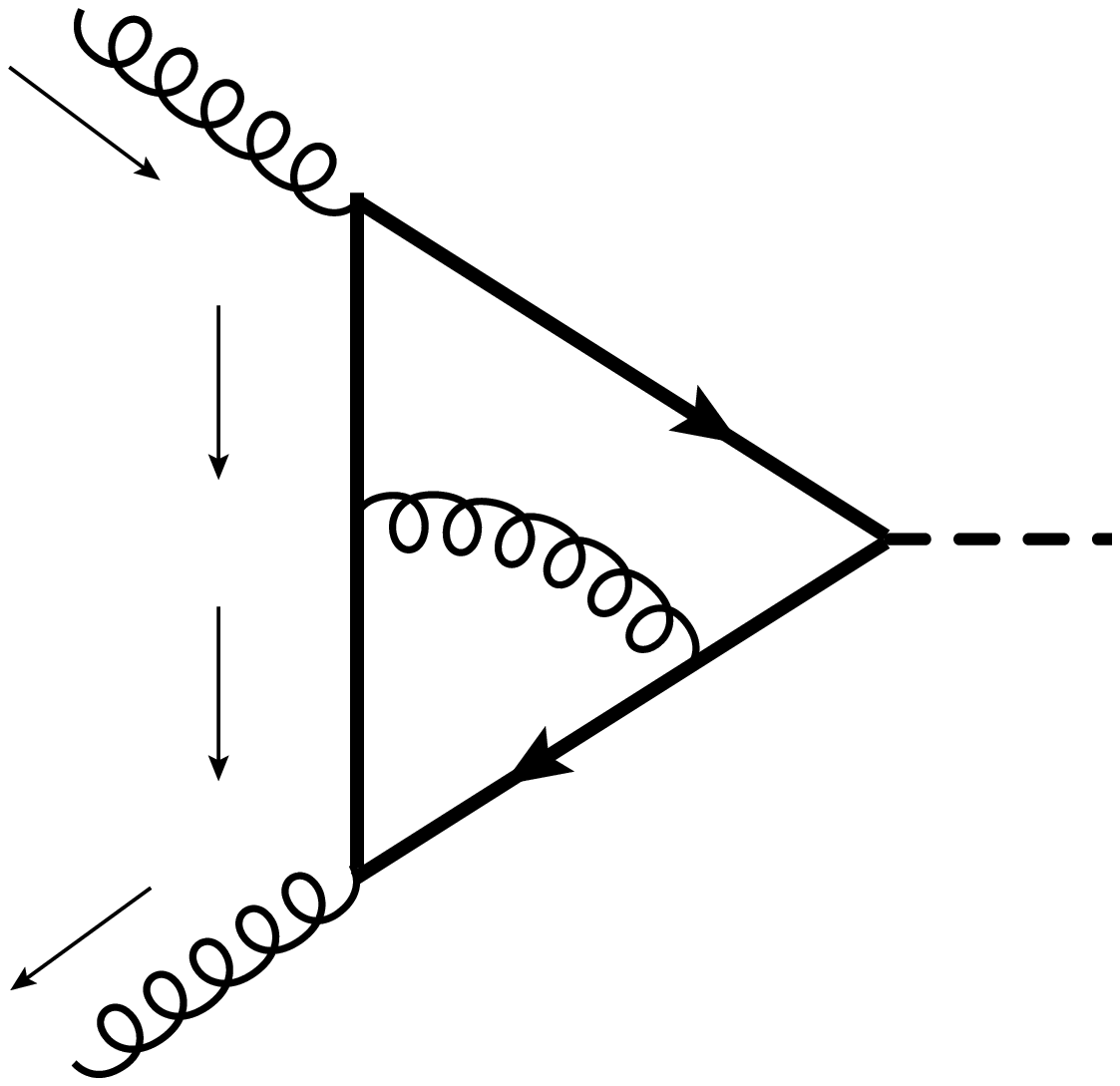}&
\hspace*{4mm}\includegraphics[width=3.5cm]{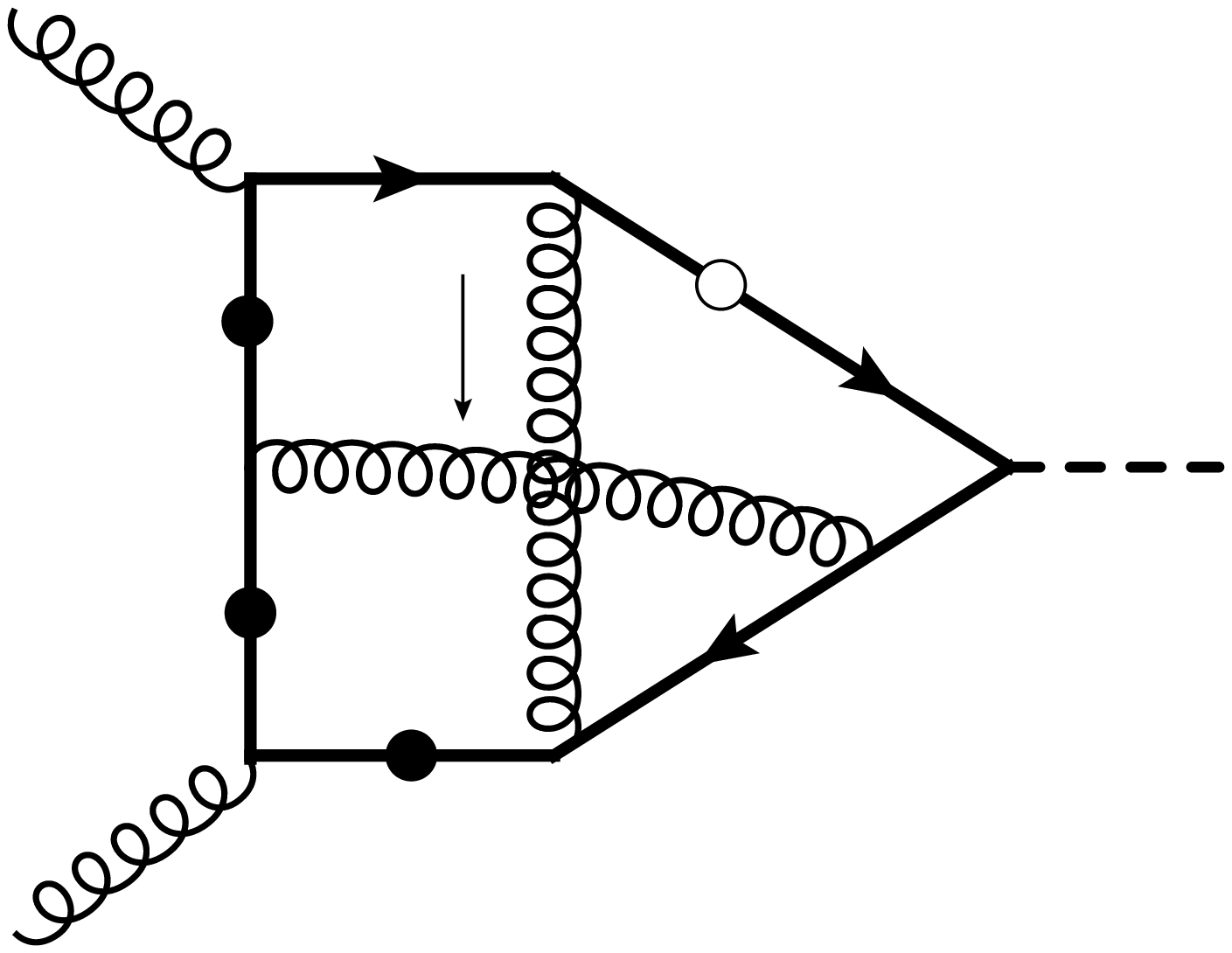}
\hspace*{-27mm}\raisebox{17.mm}{\tiny $k'_1$}\hspace*{20mm}&
\hspace*{6mm}\raisebox{1mm}{\includegraphics[width=3.5cm]{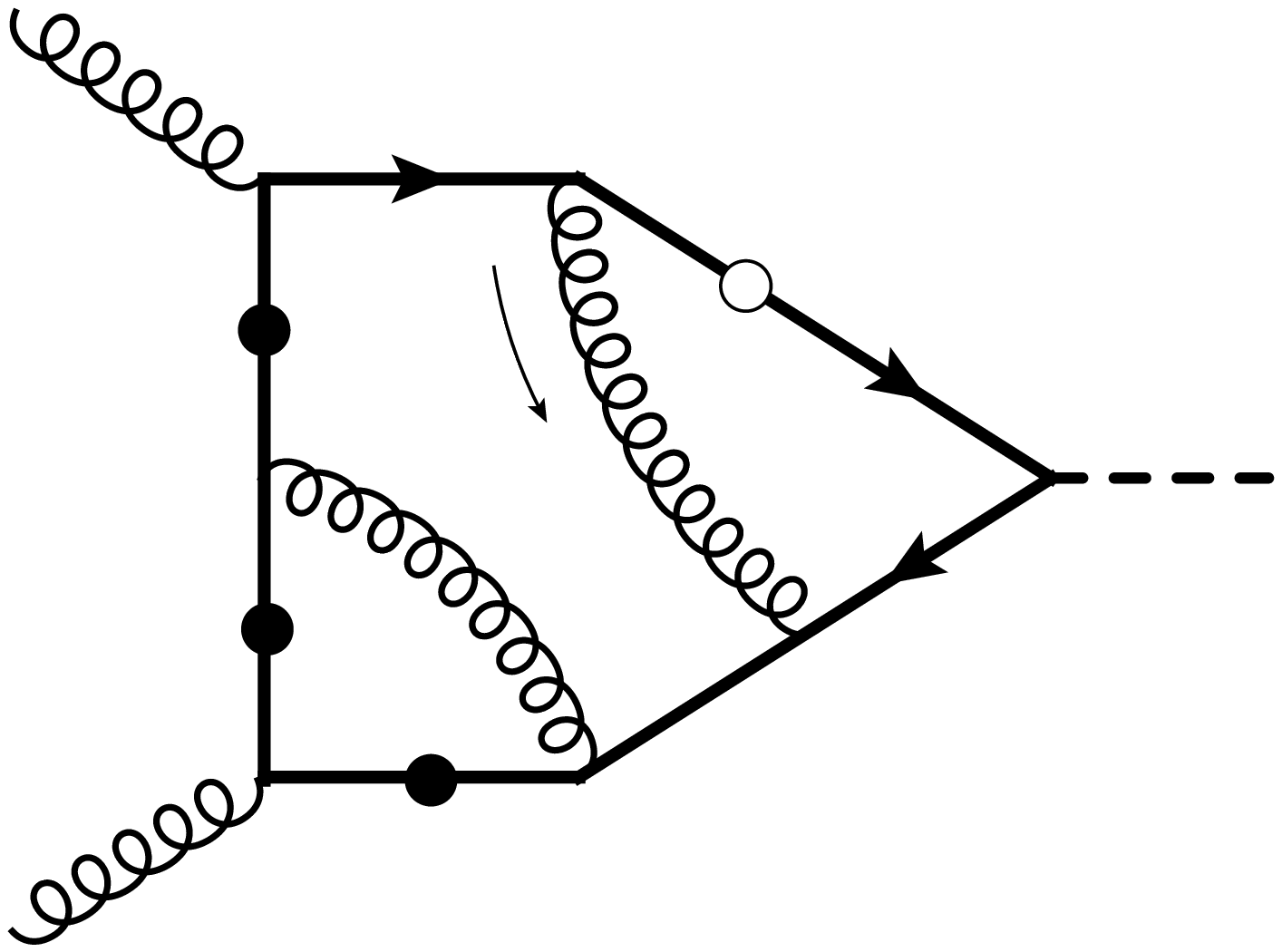}}
\hspace*{-26mm}\raisebox{17.mm}{\tiny $k_1$}\hspace*{19mm}\\
\hspace*{0mm}(a)&\hspace*{4mm}(b)&\hspace*{6mm} (c)\\
\hspace*{2mm}
\includegraphics[width=3.5cm]{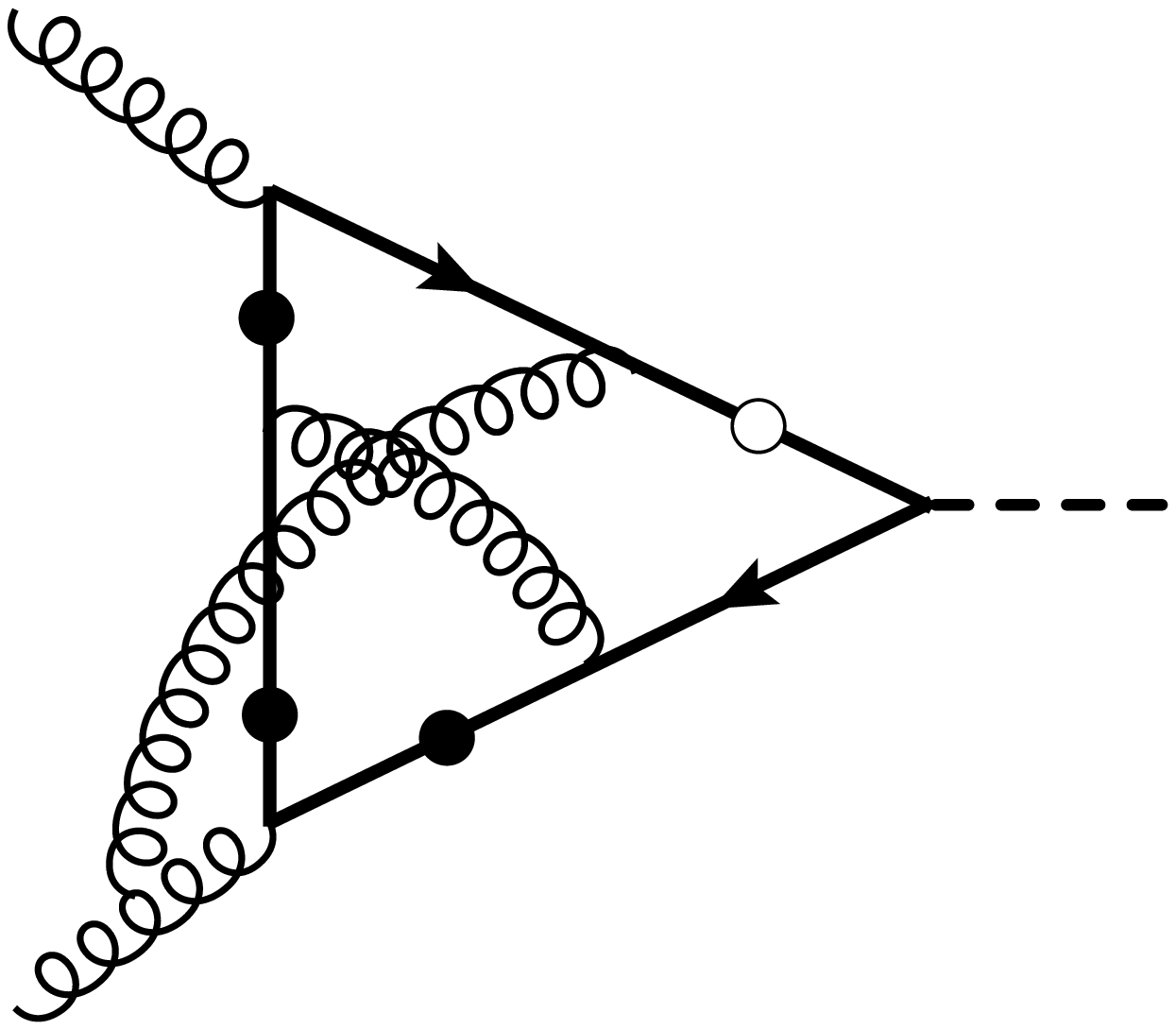}&
\hspace*{11mm}\includegraphics[width=3.5cm]{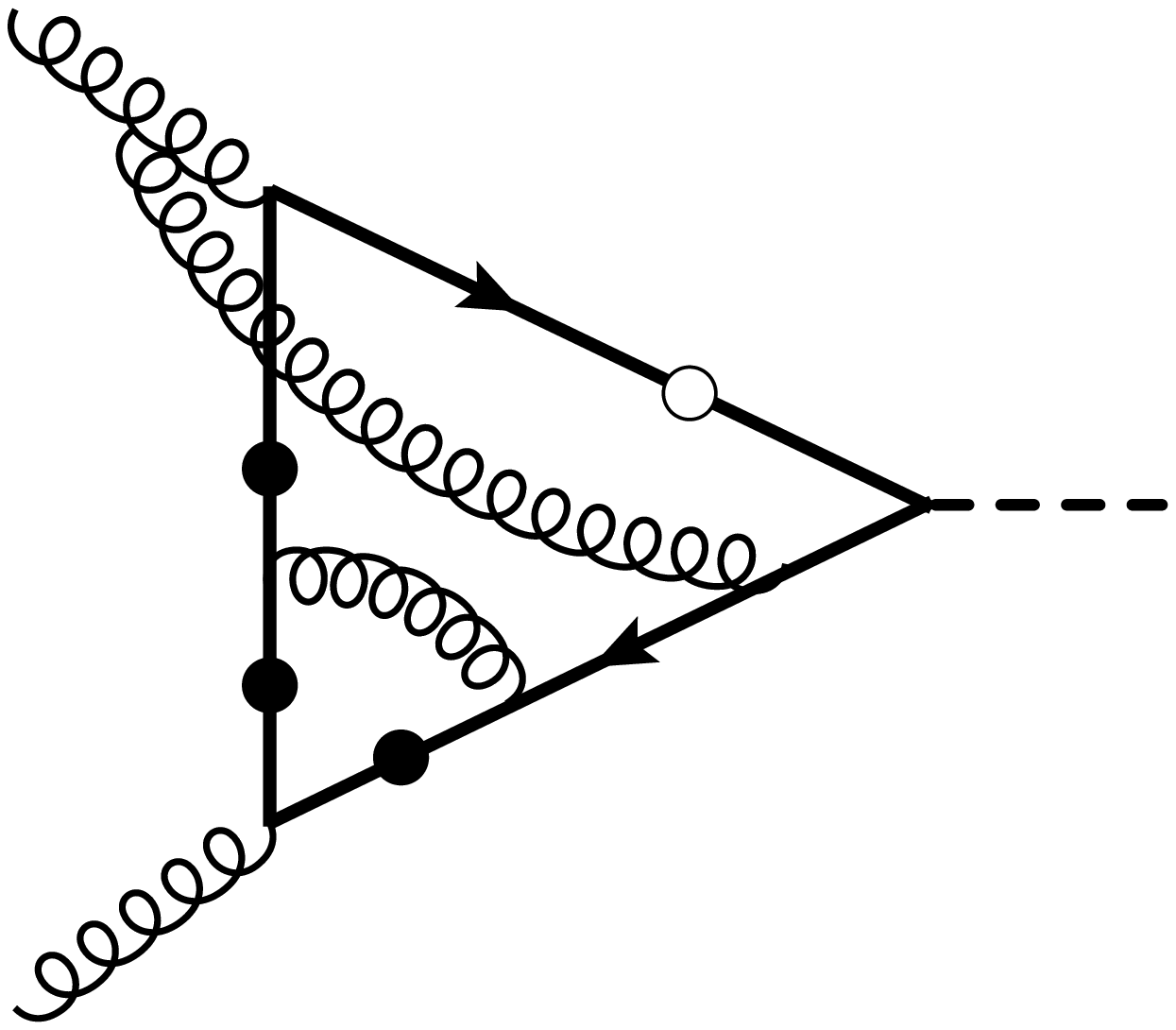}&
\hspace*{10mm}\raisebox{0mm}{\includegraphics[width=3.5cm]{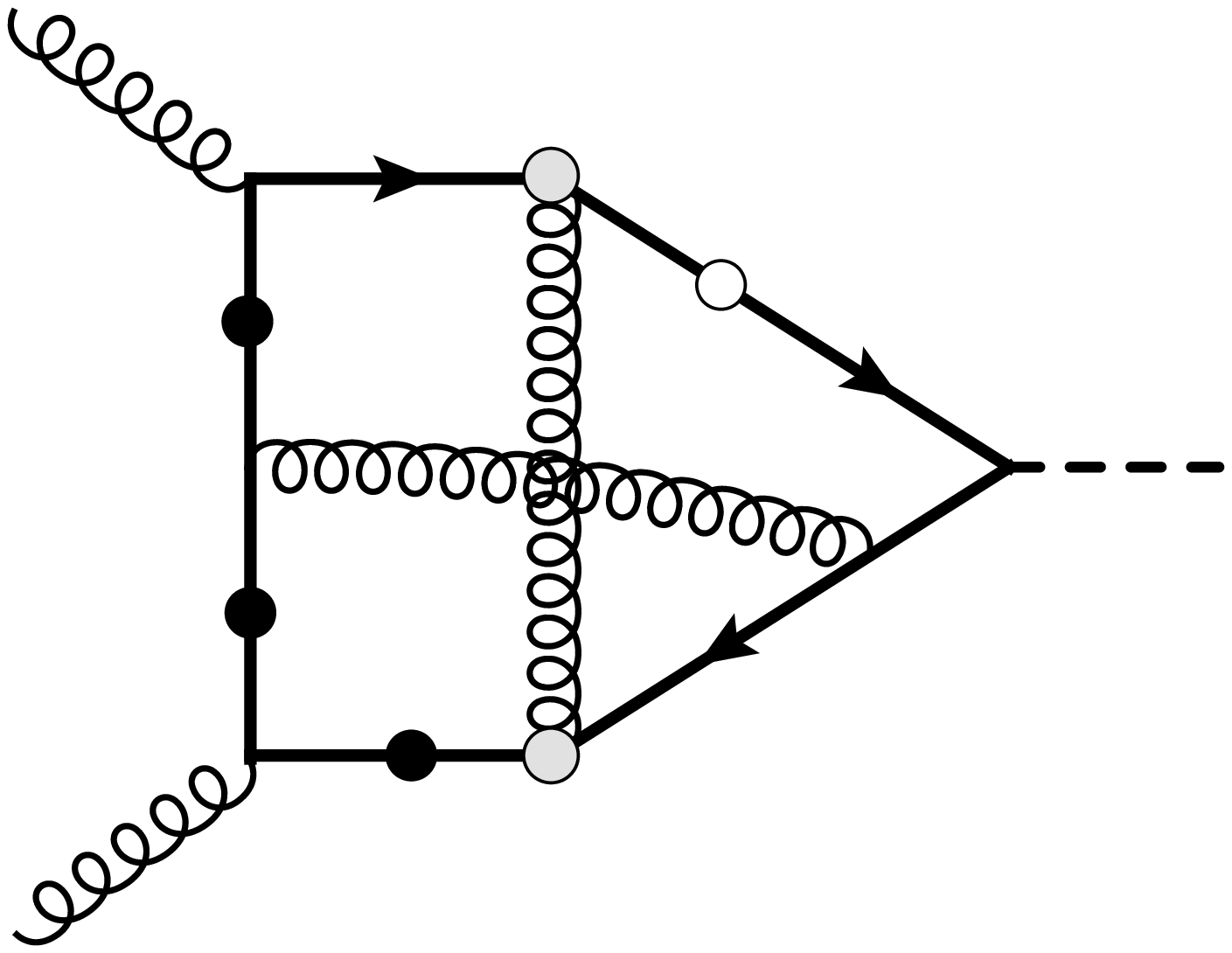}}\\
\hspace*{0mm}(d)&\hspace*{4mm}(e)&\hspace*{6.5mm} (f)\\
\end{tabular}
\end{center}
\caption{\label{fig::4}  The  Feynman diagrams representing (a)
the two-loop and  (b)-(e) the  three-loop corrections to the
$ggH$ amplitude with an additional eikonal gluon emitted by the
soft quark. (c) the Feynman diagram  with the effective soft
gluon exchange, which represents the total QCD three-loop
non-Sudakov double-logarithmic correction associated  with the
eikonal gluon emission, Eq.~(\ref{eq::3loopnfres}).}
\end{figure}

All the double-logarithmic contributions  we have considered so
far factored out into the (effective) corrections to the Higgs
boson vertex.  In three loops a new source of the
double-logarithmic corrections opens up with an additional
eikonal gluon connecting one of the eikonal and the soft  quark
lines. To trace the origin of this contributions let us
consider first the two-loop diagram in Fig.~\ref{fig::4}(a).
This diagram formally may have a double-logarithmic scaling.
Indeed,  for the virtual momentum $l$ collinear to $p_1$ and
for the soft momentum $l_1$ the gluon propagator becomes
eikonal and proportional to $1/(l_1l)$. Thus we get a standard
scalar double-logarithmic integral over $l_1$ with external
on-shell momenta $p_2$ and $l_1$ which is  proportional to the
eikonal factor $1/(p_2l)$. To get the  double-logarithmic
scaling of the integral over $l_2$ it must be canceled since
the same factor is already present in the second eikonal quark
propagator. However, due to transversal polarization of the
external gluons the relevant structure with the light-cone
component of the momentum $l$ does not appear in the tensor
decomposition of the Feynman diagram and the double-logarithmic
contribution of this type does not appear in two loops. At the
same time the relevant tensor structures do appear in three
loops. The corresponding abelian and nonabelian diagrams are
given in Figs.~\ref{fig::4}(b,c)  and (d,e), respectively. The
details of the calculation  of  the abelian contribution  are
given in Appendix~\ref{sec::appB}. Note that for the planar
topology Fig.~\ref{fig::4}(c) due to a cancellation specific to
three loops the double-logarithmic contribution vanishes.
Moreover, in the diagram Fig.~\ref{fig::4}(e) the contribution
of the soft gluon momentum coming from the three-gluon vertex
is not included. Though such a term does produce a
double-logarithmic contribution for a given diagram, it is
proportional to the momentum of the on-shell soft gluon and
after we cut the corresponding gluon line it vanishes in the
sum of the diagrams by the Ward identity. After separating the
infrared divergencies in the same way as it has been done for
the functions $f(z)$ and $g(z)$, the remaining infrared finite
double-logarithmic contribution is described by the diagram
Fig.~\ref{fig::4}(f) with the effective soft gluon exchange.
The corresponding Feynman integral is the same as for the
abelian diagram in Fig.~\ref{fig::4}(b) computed in
Appendix~\ref{sec::appB}, which gives the following
contribution to $M^{(1)}_{ggH}$
\begin{equation}
-\ln^2\!\rho\,{(C_A-C_F)\left(C_A-2C_F\right)\over 9}{x^2}\,.
\label{eq::3loopnfres}
\end{equation}
The color structure of Eq.~(\ref{eq::3loopnfres}) is quite
peculiar. As it has been previously discussed  the factor
$C_A-C_F$ accounts for the eikonal color charge variation
caused by a soft quark emission. The remaining factor
$C_A-2C_F$ reflects the change of the eikonal quark and
antiquark  state into color octet after the emission of the
eikonal gluon.

The higher-order double-logarithmic corrections of this type
are obtained by dressing the diagram in Fig.~\ref{fig::4}(a)
with multiple soft gluons.  This results in multiplication of
Eq.~(\ref{eq::3loopnfres}) by a function of the
double-logarithmic variable $j(z)= 1+\sum_{n=1}^\infty j_n
z^n$. Thus the complete  double-logarithmic approximation for
the next-to-next-to-leading power coefficient can be written as
follows
\begin{equation}
M^{(1)}_{ggH}=\ln^2\!\rho\left[-4g(z)
+\left({T_FC_F\over  45}h(z)-{(C_A-C_F)(C_A-2C_F)\over 9}j(z)
\right)x^2\right].
\label{eq::M1}
\end{equation}

\noindent
Calculation of the functions $j(z)$  requires a systematic
factorization of the soft emissions with respect to the
emission of the additional eikonal gluon. For QCD this is a
rather complicated  computational  problem  due to the soft
interaction of the eikonal gluon, which starts to contribute in
four loops. The full QCD analysis, however, goes beyond the scope
of the present paper. Instead,  we consider two complementary
limits  where such a complication is absent. First we discuss
QCD with the large number of colors  $N_c\to\infty$. In this
case the color factor of the diagram Fig.~\ref{fig::4}(a)
vanishes and the double logarithmic approximation  is entirely
determined by the function $g(z)$  where $z=N_cx/2$. In the
opposite abelian limit $C_A=0$ the gluon self-coupling is
absent but the analysis of the factorization is nevertheless
quite nontrivial, see Appendix~\ref{sec::appB}. For $C_A=0$  we
get the following integral representation of the function
$j(z)$
\begin{eqnarray}
j^{\rm ab}(z)&=&72\int_0^1{\rm d}\eta\int_{0}^{1-\eta}{\rm d}\xi
\int_{0}^{1-\xi}\!\!\!{\rm d}\eta_1\int_{0}^{1-\eta_1-\xi}\!\!\!
{\rm d}\xi_1\,\eta\xi_1 e^{2z\eta(\xi+\xi_1)}
\nonumber \\
&\times&
\left[1+{e^{-2z\eta\xi}-1\over 2}+
{e^{-2z\eta\xi}-1+2z\eta\xi\over 4z\eta\xi_1}
\right],
\label{eq::jab}
\end{eqnarray}
where $z=-C_Fx$. The first eight coefficients of the Taylor
series for $j^{\rm ab}(z)$ are listed in Table~\ref{tab::1}.
The  perturbative expansion of  Eq.~(\ref{eq::M1}) reads
\begin{eqnarray}
M^{(1)}_{ggH}&=&\ln^2\!\rho\left[-4-{2\over 3}(C_A-C_F)x
+\left({T_FC_F\over 45}-{14\over 45}C_F^2
+{23 \over 45}C_FC_A-{9\over 45}C_A^2 \right)x^2\right.
\nonumber\\
&+&c_4 x^3+\ldots \bigg],
\label{eq::M1series}
\end{eqnarray}
where the four-loop coefficient  is
$c_4=-{N_c^3/840}$
in the large-$N_c$ approximation and
\begin{equation}
c_4=-{T_FC_F^2\over 210}+{13\over 90}C_F^3
\label{eq::c4abelian}
\end{equation}
in the abelian approximation.
The series Eq.~(\ref{eq::M1series}) can be compared to the
existing fixed-order results. The two-loop term agrees  with
the expansion of the exact analytic result
\cite{Aglietti:2006tp}. The high-energy expansion of the
three-loop $ggH$ amplitude has been obtained  numerically in
Ref.~\cite{Czakon:2020vql}. Eq.~(\ref{eq::M1series})
corresponds to the following coefficient of the $L_s^6/z^2$
term in Eq.~(C.1) of \cite{Czakon:2020vql}
\begin{equation}
{1\over 23040}
\left(-{T_FC_F}+{14}C_F^2-{23}C_FC_A+9C_A^2
\right)\,,
\label{eq::C1coef}
\end{equation}
which agrees with its numerical value $0.0005738811728$. The
result Eq.~(\ref{eq::M1}) for the gluon fusion amplitude can be
transformed into the one for  the amplitude of the Higgs boson
two-photon decay by changing the color charge of the external
lines from $C_A$ to zero. This results in the replacement
$C_A-C_F\to -C_F$ in the definition of the double-logarithmic
variable $z$ and in the coefficient of
Eq.~(\ref{eq::3loopnfres}). By adopting the notations similar
to the gluon fusion case we get
\begin{equation}
M^{(1)}_{H\gamma\gamma}=\ln^2\!\rho\left[-4+{2\over 3}C_Fx
+\left({T_FC_F\over 45}-{14\over 45}C_F^2
+{C_FC_A\over 9}\right)x^2+\ldots\right].
\label{eq::M1seriesgamma}
\end{equation}
The three-loop term can be compared to the numerical result for
the high-energy expansion of the amplitude given in
Ref.~\cite{Niggetiedt:2020sbf}. It corresponds to the
coefficient
\begin{equation}
-{1\over 3840}
\left({T_FC_F}-{14}C_F^2+5{C_FC_A}\right)
\label{eq::C1coefgamma}
\end{equation}
of the $L_s^6/z^2$  term in Eq.~(C.1)  and agrees with its
numerical value $0.001099537037$. The agreement holds for the
contributions of the individual color factors
\cite{Niggetiedt:2020pc}.

Let us consider the all-order asymptotic behavior of the
${\cal O}(m_q^3)$ amplitude in the high-energy (small-mass)
limit. In the large-$N_c$ approximation it reads
\begin{equation}
M^{(1)}_{ggH}= -4\ln^2\!\rho\, g\!\left({N_cx\over 2}\right)\,,
\label{eq::M1gasymp}
\end{equation}
where
\begin{equation}
g(z)\sim \left({2\pi e^z\over z^3}\right)^{1\over 2}\,
\label{eq::gasympp}
\end{equation}
at $z\to \infty$, {\it i.e.} the amplitude is exponentially
enhanced. Note that the limit $N_c\to\infty$ is taken first and
in general may not commute  with the kinematical limit
$z\to\infty$. In the abelian approximation the relevant
asymptotic behavior of the functions in Eq.~(\ref{eq::M1}) at
$z\to -\infty$  reads
\begin{equation}
g(z)\sim -{\ln(-2z) +\gamma_E\over z}, \quad
h(z)={\cal O} (1/z^3),\quad
j^{\rm ab}(z)\sim {9\over 2 z^2}\,.
\label{eq::gasympn}
\end{equation}
Thus the coefficient  asymptotically approaches the value
$M^{(1)}_{ggH}= -\ln^2\!\rho$, {\it i.e.} the double
logarithmic corrections effectively reduce the leading-order
coefficient by factor four.

Now we can estimate the effect of the high-order ${\cal
O}(m_q^3)$ terms for the physical values of the parameters. The
relative correction to the ${\cal O}(m_q)$ amplitude is  given
by the factor
\begin{equation}
1+\rho\left[-4
+\left({T_FC_F\over  45}h(z)-{(C_A-C_F)(C_A-2C_F)\over  9}
j(z)\right){x^2\over g(z)}\right].
\label{eq::relcor}
\end{equation}
In the large-$N_c$ approximation  Eq.~(\ref{eq::relcor})
reduces to $1-4\rho$ with $\rho\approx 1.6\cdot10^{-3}$, which
amounts of approximately $0.64\%$  negative correction to the
${\cal O}(m_q)$  contribution. It does not depend on $x$ and is
the same for the gluon and photon external lines. Hence it gives a
universal all-order estimate of the next-to-next-to-leading
power corrections both for the production and decay amplitudes.

\section{Summary}
\label{sec::4}
We have studied the high-energy  asymptotic behavior of the
electromagnetic quark scattering and  the light quark loop
mediated Higgs boson production in the third order of  the
small  quark mass expansion. To our knowledge this is the first
example of the renormalization group analysis  of the
next-to-next-to-leading power amplitudes.

For the ${\cal O}(m_q^3)$ quark scattering the asymptotic
behavior is determined  by the double-logarithmic corrections
to the Pauli form factor with the structure similar to the
Dirac and scalar form factors at ${\cal O}(m_q^2)$. These
non-Sudakov double logarithms result from the eikonal color
charge nonconservation in the process with the exchange of the
soft virtual quark pair. They are described by a universal
function which shows exponential growth for the large values of
the double-logarithmic variable in QED and a logarithmic
scaling in QCD. We present for the first time
the complete  analytic asymptotic result for this function,
Eqs.~(\ref{eq::fasymQED},\ref{eq::fasymQCD}).

The double-logarithmic corrections to the ${\cal O}(m_q^3)$
Higgs boson production and  decay amplitudes  are induced by
single and triple soft quark exchanges. This is the first
example where the mass suppression of the double-logarithmic
contribution  is not entirely associated with the chirality
flip on a fermion line. Starting with three loops a new source
of the double-logarithmic corrections  opens up with an
emission of an additional virtual  eikonal gluon by the soft
quark. Our  analytic result agrees with the previous numerical
evaluation of the three-loop QCD corrections to the Higgs boson
production \cite{Czakon:2020vql} and two-photon decay
\cite{Niggetiedt:2020sbf}.   Beyond three loops the all-order
double-logarithmic asymptotic behavior of the amplitudes has
been derived in two complementary approximations. In the
large-$N_c$ limit,  which is supposed to catch the qualitative
behavior of real QCD, the structure of the double-logarithmic
corrections significantly simplifies and becomes identical to
the one of the leading ${\cal O}(m_q)$ contribution, which is
exponentially enhanced for the large values of the
double-logarithmic variable.  The opposite  abelian limit
$C_A=0$, though less phenomenologically relevant,  reveals a
more complex structure of the double-logarithmic contributions
and represents the general case for the mass-suppressed
amplitudes at the next-to-next-to-leading power.

We have also presented a quantitative estimate of the accuracy
of the high-order calculations  based on the small-mass
expansion  for the Higgs boson production  and decays mediated
by the bottom quark loop, which may become relevant with the
permanently increasing accuracy of the QCD predictions
\cite{Davies:2021zbx}. On the basis of  the double-logarithmic
analysis we conclude  that neglecting the terms  suppressed by
the mass ratio $m_b^2/m_H^2$ in such a calculation introduces a
relative error at the scale of one percent in every order of
the perturbative expansion. Our result can also be generalized
to estimate the high-order subleading top quark mass
effects on the double Higgs boson production in the high-energy
limit \cite{Davies:2018qvx,Davies:2019dfy}, where the role of the
next-to-next-to-leading power terms could be significant.

\acknowledgments
We would like to thank Marco Niggetiedt for providing  us
with the color decomposition of the three-loop result
\cite{Niggetiedt:2020sbf}. A.P.  is grateful to Vladimir
Smirnov for useful communications.  The work of T.L. is
supported in part by IHEP under Grants No.~Y9515570U1. The work
of S.M. is supported by NSERC. The work of A.P. is supported in
part by NSERC and Perimeter Institute for Theoretical Physics.

\appendix

\section{Evaluation of the function $f(\pm z)$ in the limit
$z\to\infty$}
\label{sec::appC}
It is more convenient to use an alternative integral
representation of $f(z)$
\begin{equation}
f(z)=24\int_0^1{\rm d}\eta_2
\int_{0}^{1-\eta_2}{\rm d}\xi_2\,e^{2z\eta_2\xi_2}
\int_{0}^{\eta_2}{\rm d}\eta_1\int_{0}^{\xi_2} {\rm d}\xi_1
\,e^{-2z\eta_1\xi_1}\,,
\label{eq::frep}
\end{equation}
which is equivalent to Eq.~(\ref{eq::f}). Then, the integration
over $\xi_1$ and $\eta_1$ can be done explicitly with the
result
\begin{equation}
{1\over 2z}\left(-{\rm Ei}(-2z\eta_2\xi_2)
+\ln(2z\eta_2\xi_2)+\gamma_E\right) \,.
\label{eq::intxi1eta1}
\end{equation}
For the remaining integrals we introduce new variables
$\xi_2=y\lambda^{1/2}$,  $\eta_2=\lambda^{1/2}/y$ with the
Jacobian $|J|=y$ and the integration limits
\begin{equation}
{1-(1-4\lambda)^{1/2}\over 2\lambda^{1/2}}<
y<{1+(1-4\lambda)^{1/2}\over 2\lambda^{1/2}}\,,
\qquad 0<\lambda<1/4\,.
\label{eq::lamyvar}
\end{equation}
The logarithmic integral over $y$ gives
\begin{equation}
\ln\left({1+(1-4\lambda)^{1/2}\over 1-(1-4\lambda)^{1/2}}\right)\,.
\label{eq::yint}
\end{equation}
The further analysis  depends on the sign of the argument of
$f(z)$. For $z\to\infty$ the factor
$e^{2z\eta_2\xi_2}=e^{2z\lambda}$ provides exponential
enhancement and the integral  over $\lambda$ is saturated with
the region in vicinity of the maximal value  $\lambda = 1/4$.
Thus Eq.~(\ref{eq::intxi1eta1}) can be approximated as follows
\begin{equation}
{1\over 2z}\left(\ln(z/2)+\gamma_E\right) \,,
\label{eq::posz}
\end{equation}
and Eq.~(\ref{eq::yint}) reduces to $2(1-4\lambda)^{1/2}$.
The asymptotic expansion of the resulting integral is
straightforward
\begin{equation}
\int_0^{1\over 4}2(1-4\lambda)^{1\over 2}e^{2z\lambda}{\rm d}\lambda
\sim \left(\pi e^z\over 2z^3\right)^{1\over 2}\,,
\label{eq::poslamint}
\end{equation}
which gives Eq.~(\ref{eq::fasymQED}).

For the negative value of the argument  $f(-z)$ at $z\to\infty$
the asymptotic expansion of the resulting integral is more
involved but can be performed by the standard techniques
\begin{equation}
\int_0^{1\over 4}\ln\left({1+(1-4\lambda)^{1/2}
\over 1-(1-4\lambda)^{1/2}}\right)
\left({\rm Ei}(2z\lambda)-\ln(2z\lambda)-\gamma_E\right)
e^{-2z\lambda}{\rm d}\lambda\sim
{\left(\ln{(2z)}+\gamma_E\right)^2 -\pi^2/2\over 4z},
\label{eq::neglamint}
\end{equation}
which gives Eq.~(\ref{eq::fasymQCD})

\section{Off-shell scalar form factor of massive quark}
\label{sec::appA}
We consider the scalar form factor of a quark with the off-shell
external momenta $\Delta_i=(p_i-l)^2-m_q^2\ne 0$ and
Minkowskian momentum transfer $(p_2-p_1)^2=m_H^2$. In the
double-logarithmic approximation we can set $l^2=m_q^2$ so that
$\Delta_1=-2(p_1l)=vm_H^2$, $\Delta_2=-2(p_2l)=um_H^2$, and
consider the case $|\Delta_i|\gg m_q^2$. The form factor  can
be simultaneously expanded in $\rho=m_q^2/m_H^2$ and
$\Delta_i/m_H^2$ as follows
\begin{equation}
F_S(\eta,\xi)=Z_{q}^2(\eta,\xi)\sum_{n=0}^\infty
\rho^n F^{(n)}_S(\eta,\xi)+{\cal O}(\Delta_i/m_H^2)\,,
\label{eq::FSseries}
\end{equation}
where  $Z_{q}^2(\eta,\xi)$, $F^{(n)}_S(\eta,\xi)$ are the
functions of the logarithmic variables $\eta$, $\xi$. Though
formally $|\Delta_i|\gg m_q^2$ we are not interested in the
terms  vanishing for $\Delta_i=0$ since they do not produce the
double-logarithmic corrections to the Higgs boson decay
amplitude. As in Eq.~(\ref{eq::Fiseries}) the coefficient
\begin{equation}
Z_{q}^2(\eta,\xi)=e^{-2C_Fx\eta\xi}
\label{eq::Zqetaxi}
\end{equation}
represents the usual Sudakov factor now computed for the
off-shell quarks. Thus in the double-logarithmic approximation
the leading-power coefficient is just $F^{(0)}_S(\eta,\xi)=1$.
The next-to-leading power coefficient $F^{(1)}_S(\eta,\xi)$
gets the double-logarithmic contribution from the diagrams with the
soft quark pair exchange,  Figs.~\ref{fig::2}(c,d), and has
been evaluated for the on-shell quarks in
\cite{Liu:2017vkm,Liu:2018czl}. This analysis can be extended
to the off-shell case in a straightforward way by changing the
infrared cutoff from $m_q^2$ to $\Delta_i$. The result reads
\begin{equation}
F^{(1)}_S(\eta,\xi)=8C_FT_Fx^2\int_0^\eta{\rm d}\eta_2
\int_{0}^{\xi}{\rm d}\xi_2\int_{0}^{\eta_2}{\rm d}\eta_1
\int_{0}^{\xi_2} {\rm d}\xi_1
\,e^{-2z\eta_2\xi_2}e^{2z\eta_1\xi_1}\,,
\label{eq::FSresult}
\end{equation}
where the exponential factors corresponding  to the diagrams
Fig.~\ref{fig::2}(c) and Fig.~\ref{fig::2}(d) are given
separately. The only difference of the above equation with
respect to Eqs.~(12,13) of \cite{Liu:2017vkm} is in the
integration limits over the logarithmic Sudakov variables
$\eta_i$, $\xi_i$ corresponding to each loop momenta. Since we
perform the calculation for $m_q=0$ there is no correlation
between the $\eta_i$ and  $\xi_i$ variables  unlike
Eq.~(\ref{eq::J1uv}) and the logarithmic integration intervals
in Eq.~(\ref{eq::FSresult}) are given just by ordering
these variables along the eikonal lines $\xi>\xi_2>\xi_1$,
$\eta>\eta_2>\eta_1$.

\section{Evaluation of the function $j(z)$ in abelian approximation}
\label{sec::appB}

Let us begin with the calculation of the leading three-loop
term. In the abelian approximation only two diagrams in
Figs.~\ref{fig::4}(b,c) may have double-logarithmic scaling. We
consider the nonplanar topology first. Defining the loop
momenta $l$ and $l_1$ as in Fig.~\ref{fig::4}(a) we introduce
the following Sudakov   parametrization
$l_1=u_1l+v_1p_2+{l_1}_\perp$,
$k'_1=r'_1p_1+w'_1p_2+{k'_1}_\perp$ and assume that in the
light-cone coordinates $p_1=p_1^-$, $p_2=p_2^+$. Then the
on-shell condition for the soft quark propagators requires
$uv>\rho$, $u_1v_1>\rho/u$,  and the logarithmic scaling of the
integrals over the Sudakov parameters imposes the conditions
$v<w'_1$, $uu_1<r'_1<u$. The double-logarithmic scalar integral
over $l_1$ results in the factor $1/(p_2l)=1/(p^+_2l^-)$ which
has the same structure as the  lower eikonal quark propagator
\begin{equation}
S(p_2-k'_1-l)= - {\gamma^-\over l^-}+\ldots\,,
\label{eq::p2propp}
\end{equation}
where we used the relation $k'_1{}^-\ll l^-$ valid in the
logarithmic integration region. To get the logarithmic integral
over $l^-$  one of the $1/l^-$ factors must be cancelled. By
taking into account that the real (virtual) gluons have
transversal (light-cone) polarization we find that the only
relevant tensor structure is given by the $l^-$ term in the
numerator of the upper eikonal  quark propagator
\begin{equation}
S(p_1-k'_1-l)= - {\gamma^+\over k'_1{}^+}
\left(1-{k'_1{}^-+l^-\over p_1^-}\right)+\ldots\,,
\label{eq::p1propp}
\end{equation}
as it is indicated in  Fig.~\ref{fig::4}(b). The integral over
$r'_1$ and $w'_i$ within the logarithmic limits specified above
results in the standard one-loop Sudakov correction factor
$2z\eta\xi_1$, where $z=-C_Fx$ and we introduce the logarithmic
variables $\eta_1=\ln v_1/\ln\rho$ and $\xi_1=\ln u_1/\ln\rho$ in
the same way as for the loop momentum $l$. Then the $C_A=0$
abelian double-logarithmic contribution of  the diagram
Fig.~\ref{fig::4}(b) to the coefficient $M^{(1)}_{ggH}$ reads
\begin{equation}
-8\ln^2\!\rho\,z^2\int_0^1{\rm d}\eta
\int_{0}^{1-\eta}{\rm d}\xi\int_{0}^{1-\xi}\!\!\!{\rm d}\eta_1
\int_{0}^{1-\eta_1-\xi}\!\!\!{\rm d}\xi_1\,\eta\xi_1
=-\left({\ln\!\rho\,z\over 3}\right)^2.
\label{eq::5bint}
\end{equation}
In the case of the planar diagram Fig.~\ref{fig::4}(c) the
logarithmic intervals for the Sudakov parameters of the soft
gluon momentum $k_1=r_1p_1+w_1p_2+{k_1}_\perp$  are $v<w_1$,
$u<r_1$ and the integration over $k_1$ gives the factor
$2z\eta\xi$. In contrast to the nonplanar case the required
$l^-$ term is generated in two different ways. Indeed, in the
logarithmic integration region now  $l^-\ll k_1{}^-$ and the
denominator of the  lower quark propagator can be
expanded as follows
\begin{equation}
S(p_2-k_1-l)= - {\gamma^-\over k_1{}^-}
\left(1-{l^-\over k_1{}^-}\right)+\ldots\,,
\label{eq::p2prop}
\end{equation}
while the upper quark propagator has the expansion similar
to Eq.~(\ref{eq::p1propp})
\begin{equation}
S(p_1-k_1-l)= - {\gamma^+\over k_1^+}
\left(1-{k_1^-+l^-\over p_1^-}\right)+\ldots\,.
\label{eq::p1prop}
\end{equation}
The product of  ${l^-/ k_1^-}$  term in Eq.~(\ref{eq::p2prop})
and  ${k_1^-/ p_1^-}$  term  in Eq.~(\ref{eq::p1prop})
generates the relevant tensor structure which  cancels the
contribution  from the product of the leading term in
Eq.~(\ref{eq::p2prop}) and  ${l^-/ p_1^-}$  term  in
Eq.~(\ref{eq::p1prop}).  Thus the total double-logarithmic
contribution of the planar three-loop diagram to the
coefficient $M^{(1)}_{ggH}$ vanishes  and Eq.~(\ref{eq::5bint})
with the equal contribution of the symmetric diagram
determines the abelian part of Eq.~(\ref{eq::3loopnfres}).
The above cancellation, however,  does not hold for the multiple
soft gluon exchanges.

Let us now consider the case of  $n$ soft gluons. In the abelian
approximation the  double-logarithmic corrections are generated
by the diagrams with $m'$ leading-power exchanges of the
topology Fig.~\ref{fig::4}(b) and $m=n-m'$ exchanges of the
topology Fig.~\ref{fig::4}(c), with all possible permutations
of  $n$ vertices along the upper  quark line. Let $k'_i$ and
$k_i$ be the momenta of the gluons from the first and the
second group, respectively. Each of the $n$ gluons contributes
the term $l^-$ from the numerator of the eikonal quark
propagator, as in Eqs.~(\ref{eq::p1propp},\ref{eq::p1prop}).
After the summation over all the permutations of the $n$
vertices the integrals over the $k'_i{}^+$ and  $k_i^+$
factorize.  After the (redundant) summation over $m'!m!$
permutations of vertices along the lower quark line  within
each group the integrals over the $k'_i{}^-$ and $k_i{}^-$
also factorize. Thus the $n$-loop soft contribution can  easily
be evaluated with the result
\begin{equation}
{n\over m'!m!}(2z\eta\xi_1)^{m'}(2z\eta\xi)^{m}\,.
\label{eq::nloopl}
\end{equation}
The analysis of the $l^-$ terms originating from the
denominators of the eikonal quark propagators, as in
Eq.~(\ref{eq::p2prop}), is more subtle. These  terms are
generated by the soft gluons from the second group only and
come from the expansion of the following expression
\begin{equation}
{f_1k^-_1+\ldots+f_mk^-_m\over (k^-_1+l^-)
\ldots (k^-_1+\ldots+k^-_m+l^-)}\,,
\label{eq::trick1}
\end{equation}
where the gluon momenta  are enumerated from the eikonal gluon
to the Higgs boson vertex and $f_i$ is the number of the
eikonal propagators carrying the momentum $k_i$ on the upper
quark line for a given  diagram. The numerator of
Eq.~(\ref{eq::trick1}) can be rewritten as follows
\begin{eqnarray}
&&f_m(k^-_1+\ldots+k^-_m+l^-)
\nonumber\\
&&+(f_{m-1}-f_m)(k^-_1+\ldots+k^-_{m-1}+l^-)+
\ldots
\nonumber\\
&&+(f_1-f_2)(k^-_1+l^-)-f_1l^-\,.
\label{eq::trick2}
\end{eqnarray}
Note that every term in Eq.~(\ref{eq::trick2})  except the last
one cancels one of the eikonal quark propagators in
Eq.~(\ref{eq::trick1}) removing the double-logarithmic scaling
of the integrand. Thus we have to consider only the
contribution of last term corresponding to the soft gluon
emitted next to the eikonal gluon vertex and the total result
is obtained by summing up the coefficients $f_1$ over the
diagrams with all possible permutations of the remaining
vertices. It is convenient to perform the double-logarithmic
integration over $k'_i$ and $k_i$ first. Since in the
logarithmic region the Sudakov parameters are ordered along the
eikonal lines, for every diagram the $n$-fold integral over
$w'_i$ and $w_i$ gives $\eta^n/n!$, the $m'$-fold integral over
$r'_i$ gives $\xi_1^{m'}/m'!$, and the $m$-fold integral over
$r_i$ gives $\xi^{m}/m!$. This combines into the common
$n$-loop factor
\begin{equation}
{(2z\eta\xi_1)^{m'}(2z\eta\xi)^{m}\over n!m'!m!}\,,
\label{eq::nloopint}
\end{equation}
Since $f_1$ does not depend on the routing of the other loop
momenta we can perform summation over the  permutations within
the groups of $m'$ and $m-1$ remaining vertices on the lower
quark line, which results in the factor $m'!(m-1)!$ for $m>0$.
Now let $j'$ and $j$ be the numbers of the vertex with the soft
momentum $k_1$  in a  sequence of all $n$ vertices and in a
sequence of $m$ vertices of the second group on the upper quark
line, respectively, counted from the Higgs boson  vertex. Then
for a given diagram $f_1=j'$ and the sum over all the diagrams
gives
\begin{equation}
\sum_{j=1}^{m}\sum_{j'=1}^{j+m'}{(j'-1)!\over (j-1)!(j'-j)!}
{(n-j')!\over (m'+j-j')!(m-j)!}j'={n!\over m!m'!}{(n+1)m\over 2}\,,
\label{eq::combfact}
\end{equation}
where the combinatorial factor corresponds to the number of
ways to arrange $m'$ ordered vertices from the first group and
$m-1$ ordered vertices from the second group for a given $j'$
and $j$. Bringing all the factors together we get
\begin{equation}
-{n+1\over 2m'!m!}(2z\eta\xi_1)^{m'}(2z\eta\xi)^{m}\,,
\label{eq::nloopk}
\end{equation}
which after adding the contribution Eq.~(\ref{eq::nloopl})
gives the total result for $m>0$
\begin{equation}
{m+m'-1\over 2m'!m!}(2z\eta\xi_1)^{m'}(2z\eta\xi)^{m}\,.
\label{eq::nlooptotm}
\end{equation}
The  $m=0$ result can be obtained directly from
Eq.~(\ref{eq::nloopl}) and reads
\begin{equation}
{(2z\eta\xi_1)^{m'}\over (m'-1)!}\,.
\label{eq::nlooptot0}
\end{equation}
The dependence on $m$ and $m'$ in Eq.~(\ref{eq::nlooptotm})
factorizes  and the summation over the number of soft gluons
in each group can be directly performed
\begin{eqnarray}
&&\sum_{m'=0}^\infty {(2z\eta\xi_1)^{m'}\over m'!}
\left[\sum_{m=0}^\infty{m+m'-1\over 2m!}
(2z\eta\xi)^{m}+{m'+1\over 2}\right]
\nonumber\\
&&={e^{2z\eta(\xi+\xi_1)}\over 2}
\left[\left(e^{-2z\eta\xi}-1+{2z\eta\xi}\right)
+(2z\eta\xi_1)\left(e^{-2z\eta\xi}+1\right)\right],
\label{eq::mmpsum}
\end{eqnarray}
where the second term in the first line provides the correct
$m=0$ contribution, Eq.~(\ref{eq::nlooptot0}). After factoring
out  the leading soft gluon contribution $2z\eta\xi_1$  we
obtain the integrand of Eq.~(\ref{eq::jab}).  Note  that the
contributions of the first and the second group of the soft
gluons completely factorize and exponentiate in the final
result.


\begin{thebibliography}{99}

\bibitem{Sudakov:1954sw}
  V.~V.~Sudakov,
  Sov.\ Phys.\ JETP {\bf 3}, 65 (1956)
  [Zh.\ Eksp.\ Teor.\ Fiz.\  {\bf 30}, 87 (1956)].


\bibitem{Frenkel:1976bj}
  J.~Frenkel and J.~C.~Taylor,
  Nucl.\ Phys.\ B {\bf 116}, 185 (1976).

\bibitem{Smilga:1979uj}
  A.~V.~Smilga,
  Nucl.\ Phys.\ B {\bf 161}, 449 (1979).

\bibitem{Mueller:1979ih}
  A.~H.~Mueller,
  Phys.\ Rev.\ D {\bf 20}, 2037 (1979).

\bibitem{Collins:1980ih}
  J.~C.~Collins,
  Phys.\ Rev.\ D {\bf 22}, 1478 (1980).

\bibitem{Sen:1981sd}
  A.~Sen,
  Phys.\ Rev.\ D {\bf 24}, 3281 (1981).

\bibitem{Sterman:1986aj}
  G.~F.~Sterman,
  Nucl.\ Phys.\ B {\bf 281}, 310 (1987).

\bibitem{Korchemsky:1988pn}
  G.~P.~Korchemsky,
  Phys.\ Lett.\ B {\bf 217}, 330 (1989).

\bibitem{Korchemsky:1988hd}
  G.~P.~Korchemsky,
  Phys.\ Lett.\ B {\bf 220}, 629 (1989).



\bibitem{Kuhn:1999nn}
  J.~H.~Kuhn, A.~A.~Penin and V.~A.~Smirnov,
  Eur.\ Phys.\ J.\ C {\bf 17}, 97 (2000).


\bibitem{Kuhn:2001hz}
  J.~H.~Kuhn, S.~Moch, A.~A.~Penin and V.~A.~Smirnov,
  Nucl.\ Phys.\ B {\bf 616}, 286 (2001),
  Erratum: [Nucl.\ Phys.\ B {\bf 648}, 455 (2003)].

\bibitem{Feucht:2004rp}
  B.~Feucht, J.~H.~Kuhn, A.~A.~Penin and V.~A.~Smirnov,
  Phys.\ Rev.\ Lett.\  {\bf 93}, 101802 (2004).


\bibitem{Jantzen:2005az}
  B.~Jantzen, J.~H.~K\"uhn, A.~A.~Penin and V.~A.~Smirnov,
  Nucl.\ Phys.\ B {\bf 731}, 188 (2005).


\bibitem{Penin:2005kf}
  A.~A.~Penin,
  Phys.\ Rev.\ Lett.\  {\bf 95}, 010408 (2005).

\bibitem{Penin:2005eh}
  A.~A.~Penin,
  Nucl.\ Phys.\ B {\bf 734},  185 (2006).


\bibitem{Bonciani:2007eh}
  R.~Bonciani, A.~Ferroglia, and A.~A.~Penin,
  Phys.\ Rev.\ Lett.\  {\bf 100},  131601 (2008).

\bibitem{Bonciani:2008ep}
  R.~Bonciani, A.~Ferroglia, and A.~A.~Penin,
  JHEP {\bf 0802},  080 (2008).

\bibitem{Kuhn:2007ca}
  J.~H.~K\"uhn, F.~Metzler and A.~A.~Penin,
  Nucl.\ Phys.\ B {\bf 795}, 277 (2008).

\bibitem{Kuhn:2011mh}
  J.~H.~K\"uhn, F.~Metzler, A.~A.~Penin, and S.~Uccirati,
  JHEP {\bf 1106}, 143 (2011).


\bibitem{Penin:2011aa}
  A.~A.~Penin and G.~Ryan,
  JHEP {\bf 1111},  081 (2011).

\bibitem{Gorshkov:1966ht}
  V.~G.~Gorshkov, V.~N.~Gribov, L.~N.~Lipatov and G.~V.~Frolov,
  Sov.\ J.\ Nucl.\ Phys.\  {\bf 6}, 95 (1968)
  [Yad.\ Fiz.\  {\bf 6}, 129 (1967)].


\bibitem{Kotsky:1997rq}
  M.~I.~Kotsky and O.~I.~Yakovlev,
  Phys.\ Lett.\ B {\bf 418}, 335 (1998).


\bibitem{Penin:2014msa}
  A.~A.~Penin,
  Phys.\ Lett.\ B {\bf 745}, 69 (2015), Erratum: [Phys.\ Lett.\ B {\bf 771}, 633
(2017)].


\bibitem{Melnikov:2016emg}
  K.~Melnikov and A.~Penin,
  JHEP {\bf 1605}, 172 (2016).



\bibitem{Penin:2016wiw}
  A.~A.~Penin and N.~Zerf,
  Phys.\ Lett.\ B {\bf 760}, 816 (2016),  Erratum: [Phys.\ Lett.\ B {\bf 771},
637 (2017)].


\bibitem{Liu:2017axv}
  T.~Liu, A.~A.~Penin and N.~Zerf,
  Phys.\ Lett.\ B {\bf 771}, 492 (2017).



\bibitem{Liu:2017vkm}
  T.~Liu and A.~A.~Penin,
  Phys.\ Rev.\ Lett.\  {\bf 119},  262001 (2017).



\bibitem{Liu:2018czl}
  T.~Liu and A.~Penin,
  JHEP {\bf 1811}, 158 (2018).



\bibitem{Liu:2019oav}
  Z.~L.~Liu and M.~Neubert,
  JHEP {\bf 2004}, 033 (2020).


\bibitem{Wang:2019mym}
  J.~Wang,
  arXiv:1912.09920 [hep-ph].


\bibitem{Anastasiou:2020vkr}
  C.~Anastasiou and A.~Penin,
  JHEP {\bf 2007}, 195 (2020).


\bibitem{Liu:2020tzd}
  Z.~L.~Liu, B.~Mecaj, M.~Neubert and X.~Wang,
  Phys.\ Rev.\ D {\bf 104} (2021) 014004.


\bibitem{Liu:2020wbn}
  Z.~L.~Liu, B.~Mecaj, M.~Neubert and X.~Wang,
  JHEP {\bf 2101}, 077 (2021).


\bibitem{Ferroglia:2009ep}
  A.~Ferroglia, M.~Neubert, B.~D.~Pecjak and L.~L.~Yang,
  Phys.\ Rev.\ Lett.\  {\bf 103} (2009) 201601.


\bibitem{Laenen:2010uz}
  E.~Laenen, L.~Magnea, G.~Stavenga and C.~D.~White,
  JHEP {\bf 1101}, 141 (2011).


\bibitem{Becher:2013iya}
  T.~Becher and G.~Bell,
  Phys.\ Rev.\ Lett.\  {\bf 112}, 182002 (2014).

\bibitem{deFlorian:2014vta}
  D.~de Florian, J.~Mazzitelli, S.~Moch and A.~Vogt,
  JHEP {\bf 1410}, 176 (2014).

\bibitem{Anastasiou:2014lda}
  C.~Anastasiou, C.~Duhr, F.~Dulat, E.~Furlan, T.~Gehrmann, F.~Herzog and
B.~Mistlberger,
  JHEP {\bf 1503}, 091 (2015)



\bibitem{Boughezal:2018mvf}
  R.~Boughezal, A.~Isgr\'o and F.~Petriello,
  Phys.\ Rev.\ D {\bf 97},  076006 (2018).

\bibitem{Bruser:2018jnc}
  R.~Br\"user, S.~Caron-Huot and J.~M.~Henn,
  JHEP {\bf 1804}, 047 (2018).

  \bibitem{Moult:2018jjd}
  I.~Moult, I.~W.~Stewart, G.~Vita and H.~X.~Zhu,
  JHEP {\bf 1808}, 013 (2018).

\bibitem{Ebert:2018lzn}
  M.~A.~Ebert, I.~Moult, I.~W.~Stewart, F.~J.~Tackmann, G.~Vita and H.~X.~Zhu,
  JHEP {\bf 1812}, 084 (2018).


\bibitem{Beneke:2018gvs}
  M.~Beneke, A.~Broggio, M.~Garny, S.~Jaskiewicz, R.~Szafron, L.~Vernazza and J.~Wang,
  JHEP {\bf 1903}, 043 (2019).

\bibitem{Engel:2018fsb}
  T.~Engel, C.~Gnendiger, A.~Signer and Y.~Ulrich,
  JHEP {\bf 1902}, 118 (2019).

\bibitem{Ebert:2018gsn}
  M.~A.~Ebert, I.~Moult, I.~W.~Stewart, F.~J.~Tackmann, G.~Vita and H.~X.~Zhu,
  JHEP {\bf 1904}, 123 (2019).

\bibitem{Penin:2019xql}
  A.~A.~Penin,
  JHEP {\bf 2004}, 156 (2020).

\bibitem{Beneke:2019mua}
  M.~Beneke, M.~Garny, S.~Jaskiewicz, R.~Szafron, L.~Vernazza and J.~Wang,
  JHEP {\bf 2001}, 094 (2020).


\bibitem{Czakon:2020vql}
  M.~L.~Czakon and M.~Niggetiedt,
  JHEP {\bf 2005}, 149 (2020).

\bibitem{Niggetiedt:2020sbf}
  M.~Niggetiedt,
  JHEP {\bf 2104}, 196 (2021).


\bibitem{Melnikov:2016qoc}
  K.~Melnikov, L.~Tancredi and C.~Wever,
  JHEP {\bf 1611}, 104 (2016).

\bibitem{Lindert:2017pky}
  J.~M.~Lindert, K.~Melnikov, L.~Tancredi and C.~Wever,
  Phys.\ Rev.\ Lett.\  {\bf 118}, no. 25, 252002 (2017).

\bibitem{Beneke:1997zp}
  M.~Beneke and V.~A.~Smirnov,
  Nucl.\ Phys.\ B {\bf 522}, 321 (1998).

\bibitem{Smirnov:1997gx}
  V.~A.~Smirnov,
  Phys.\ Lett.\ B {\bf 404}, 101 (1997).

\bibitem{Smirnov:2002pj}
  V.~A.~Smirnov,
  {\it Applied asymptotic expansions in momenta and masses},
  Springer Tracts Mod.\ Phys.\  {\bf 177 } (2002)  1.


\bibitem{Bernreuther:2004ih}
  W.~Bernreuther, R.~Bonciani, T.~Gehrmann, R.~Heinesch, T.~Leineweber,
  P.~Mastrolia and E.~Remiddi,
  Nucl.\ Phys.\ B {\bf 706}, 245 (2005).


\bibitem{Aglietti:2006tp}
  U.~Aglietti, R.~Bonciani, G.~Degrassi and A.~Vicini,
  JHEP {\bf 0701}, 021 (2007).

\bibitem{Niggetiedt:2020pc}
  M.~Niggetiedt, private communication.

\bibitem{Davies:2021zbx}
J.~Davies and F.~Herren,
Phys. Rev. D \textbf{104}, 053010 (2021).

\bibitem{Davies:2018qvx}
J.~Davies, G.~Mishima, M.~Steinhauser and D.~Wellmann,
JHEP \textbf{01}, 176 (2019).

\bibitem{Davies:2019dfy}
J.~Davies, G.~Heinrich, S.~P.~Jones, M.~Kerner, G.~Mishima, M.~Steinhauser and D.~Wellmann,
JHEP \textbf{11}, 024 (2019).



\end{thebibliography}
\end{document}